\documentclass{article}

\usepackage[english]{babel}
\usepackage[
    style=authoryear-ibid, 
    backend=biber, 
    maxcitenames=3, % If three authors,  in-text citations before "et al."
    maxbibnames=10  % Show all authors in the bibliography (adjust as needed)
]{biblatex}

\DeclareNameAlias{sortname}{family-given}
\renewbibmacro{in:}{} % This tells biblatex not to print "In:" before the book title.

\usepackage{caption}
\usepackage{subcaption}
\usepackage{siunitx}
\usepackage{lipsum}% Just for this example

\usepackage[table,xcdraw]{xcolor}
\usepackage[a4paper,top=2cm,bottom=2cm,left=3cm,right=3cm,marginparwidth=1.75cm]{geometry}

\usepackage{amsmath}
\usepackage{hyperref}
\usepackage{url}
\usepackage{graphicx}
\usepackage{floatrow}
\usepackage{authblk}
\usepackage{booktabs} % for table lines

\usepackage{csquotes}% Recommended

\usepackage[T1]{fontenc}
\usepackage{lipsum}

\title{How liveable are London's 15-minute neighbourhoods? Exploring liveability profiles and active travel patterns in London}

\author[1]{Philyoung Jeong}
\author[1,2]{Ivann Schlosser}
\author[3]{Alexei Poliakov}
\author[1]{Elsa Arcaute}

\affil[1]{Centre for Advanced Spatial Analysis, University College London, London, W1T 4TJ, UK}
\affil[2]{Oxford Programme for Sustainable Infrastructure Systems (OPSIS), University of Oxford, OX1 3AN, UK }
\affil[3]{Locomizer Ltd.}

\begin{document}
\maketitle
\normalsize
\begin{abstract}
Healthy and liveable neighbourhoods have increasingly been recognised as essential components of sustainable urban development. Yet, ambiguity surrounding their definition and constituent elements presents challenges in understanding and evaluating neighbourhood profiles, highlighting the need for a more detailed and systematic assessment. This research develops a composite Liveability Index for Greater London based on metrics related to the proximity, density, and diversity of POIs, along with population density, and investigates how neighbourhood liveability relates to active travel behaviour. The Index is based on the principles of the 15-minute city paradigm \parencite{moreno2021introducing} and developed in line with OECD guidelines for composite indicators \parencite{joint2008handbook}. The Index revealed distinct spatial patterns of neighbourhood liveability, with high liveability neighbourhoods predominantly clustered in Inner London. Decomposing the Index provided further insights into the strengths and weaknesses of each neighbourhood. Footfall modelling using ordinary least squares (OLS) and geographically weighted regression (GWR) indicates a generally positive relationship between liveability and footfall, with spatial variation in the strength of this association. This research offers a new perspective on conceptualising and measuring liveability, demonstrating its role as an urban attractor that fosters social interaction and active engagement.

\end{abstract}
%\newpage

%%%%%%%%%%%%%%%%%%%%%%%%%%%%%%%%%%%%%%%%%%%%%%%%%%%%%%%%%%%%%%%%
\section{Introduction}

With rapid urbanisation, cities worldwide confront multifaceted challenges that transcend traditional boundaries. By 2050, seven out of ten individuals will reside in urban areas \parencite{un-habitatWorldCitiesReport2022}, increasing resource demand, straining infrastructure, and impacting quality of life. Cities already consume approximately two-thirds of global energy and produce over 70\% of carbon emissions \parencite{worldbankUrbanDevelopment2023}. The Intergovernmental Panel on Climate Change (IPCC) emphasises their critical role in achieving net-zero emissions, calling for integrated planning across physical, natural, and social infrastructure \parencite{ipccClimateChange20232023}. Addressing these challenges requires a fundamental reimagining of how cities function and operate.

The pandemic has amplified the urgency for urban transformation by exposing vulnerabilities in modern cities. Movement restrictions confined people to their immediate neighbourhoods, exposing limited access to essential urban services and revealing weaknesses in the resilience of current urban systems. In response, this spurred localised initiatives aimed at creating more inclusive and liveable neighbourhoods. For instance, the Netherlands’ micro-markets and Austria’s maze-like parks demonstrate how neighbourhood-scale strategies can foster healthier, more sustainable lifestyles while accommodating social distancing \parencite{daviesHowCOVID19Could2020}. Similarly, Moreno’s 15-minute city paradigm outlines essential principles for proximity-based urban planning, highlighting the potential of localised approaches to achieve sustainability \parencite{moreno2021introducing}. These examples collectively underscore the transformative potential of neighbourhoods in tackling climate challenges and enhancing residents’ well-being.

A neighbourhood serves as a small-scale enclave within a city, characterised by its distinct identity and functionality. Whilst its definition may vary across disciplines, this study draws on Jane Jacobs’ vision of diverse and dynamic neighbourhoods that prioritise their residents — that is, neighbourhoods designed to support people’s daily activities, social interactions, and overall quality of life \parencite{jacobs1961death}. This perspective aligns with the contemporary shift in urban planning towards more human-scale approaches \parencite{gehl2013cities}.

Driven by the growing emphasis on local living and policy efforts aimed at creating people-centred communities, neighbourhood-level liveability assessment has gained timely prominence. However, a lack of clarity around the core components of liveability and the limited availability of granular assessment tools pose significant challenges to understanding and improving liveability at the neighbourhood scale. This research addresses this shortfall by developing a Liveability Index to evaluate the liveability profiles of Greater London's neighbourhoods and to examine their relationship with active travel patterns. Grounded on the principles of the 15-minute city concept, the Index encapsulates liveability through three spatial dimensions: proximity, diversity, and population density. These dimensions are closely linked to the spatial and functional characteristics of daily urban life and can be consistently mapped and measured, unlike other liveability aspects such as social or environmental factors, which are often less spatially explicit and more difficult to quantify. We show that these three dimensions suffice to capture the social attraction factors by leveraging footfall data. In addition, by allowing users to zoom into the neighbourhood level, the index offers nuanced insights into local liveability strengths and deficits that are often obscured in broader-scale assessments. Through this assessment, this study aims to provide actionable insights for developing tailored policies that address underperforming aspects and enhance neighbourhood liveability.

\section{Background}

\subsection{Shift to human-scale urban planning: the 15-minute city paradigm}
The rise of affordable cars led to the development of car-centred urban layouts, raising environmental concerns and social exclusion. In response, \textcite{jacobs1961death} championed diverse, walkable, mixed-use neighbourhoods through a resident-inclusive approach, while \textcite{whyte1980social} highlighted the value of small-scale, pedestrian-friendly areas that encourage social interactions and diverse activities. Their ideas significantly influenced New Urbanism – a movement calling for reimagining city structures by incorporating elements that prioritise community, sustainability and quality of life \parencite{congress2000charter}. 

The Covid-19 pandemic intensified the need for human-centred urban planning, exposing socio-economic disparities and the exclusionary nature of modern city designs. For example, movement restrictions confined individuals to their neighbourhoods, limiting access to essential services and disproportionately affecting marginalised populations. These challenges prompted urban planners to re-evaluate neighbourhoods at a micro scale and prioritise the creation of more liveable, inclusive spaces. Moreno’s 15-minute city concept envisions neighbourhoods as self-sufficient units where residents meet their daily needs within their activity space \parencite{pozoukidou202115}. It advocates for optimal population density and easy access to six crucial urban services: living, commerce, working, education, entertainment and healthcare \parencite{moreno2021introducing}. By reducing car dependency, this model helps cities curb carbon emissions, revitalise local economies, and attract diverse amenities, ultimately enhancing residents’ well-being \parencite{allam202215}. Beyond proximity, the paradigm emphasises density, diversity and digitalisation – where density refers to population concentration, diversity includes cultural variety and mixed land-use, and digitalisation leverages advanced technology and data to facilitate resident engagement and implementation \parencite{moreno2021introducing}. The 15-minute city concept aligns with major climate agendas, including the Paris Agreement and Sustainable Development Goal 11 \parencite{allam202215}, and its application in Paris has gained international recognition as an exemplary approach for post-pandemic urban revitalisation \parencite{pozoukidou202115, gongadzeParisVision15Minute2023}.

However, the 15-minute city concept is not without criticism. \textcite{glaeser202115} argues that it fragments cities into isolated enclaves, hindering city-wide connectivity and limiting broader opportunities. This raises concerns about balancing self-sufficient neighbourhoods with urban cohesion on a larger scale. \textcite{osullivanWhere15MinuteCity2021} cautions that it may exacerbate racial and socio-economic disparities and accelerate gentrification. Using GPS data, \textcite{abbiasov15minuteCityQuantified2024} identified a link between 15-minute city land-use and socio-economic segregation, showing that lower-income populations experience higher levels of segregation because opportunities for cross-income interactions are generally concentrated outside their neighbourhoods. These critiques underscore the need to consider the complex interplay between social and economic dynamics, as they shape distinct characteristics of urban regions. Neglecting these factors could inadvertently exacerbate existing inequalities and deepen social polarisation.

\subsection{Defining neighbourhoods}
The concept of a ‘neighbourhood’ in urban landscapes is intricate and interpreted from various perspectives. \textcite{sheppardThinkingGeographicallyGlobalizing2015} explores this complexity through epistemological lenses - positivism, idealism, and structuralism - highlighting the multifaceted nature of our comprehension of urban space. This suggests that there is no single, universal way to define a neighbourhood, and that multiple perspectives are necessary to capture its full complexity.

% neighbourhoods as mixed/diverse spaces
From a built environment perspective, neighbourhoods are often conceptualised as clusters of amenities, where their presence and spatial arrangement shape identity and functionality. For example, \textcite{doi:10.1177/0739456X14550401} examines spatial patterns of amenity clustering, revealing how retail and food establishments coalesce within neighbourhoods. These insights are extended further by employing a proximity-based clustering approach to identify neighbourhoods \parencite{hidalgoAmenityMixUrban2020}, mapping amenities to local peaks of accessibility. Building upon this, \textcite{ivann_2025_16913800} develops a clustering-based method to infer Greater London’s neighbourhoods using network proximity to a local mix of urban POIs. Their finding reveals that the POI-based neighbourhoods align closely with the significant activity spaces of residents' mobility patterns, validating this approach to define neighbourhoods.

% Data
Advancements in mobility data offer a more nuanced understanding of how people use space and how this, in turn, shapes neighbourhoods. \textcite{gonzalezUnderstandingIndividualHuman2008} reveal that human mobility is not random but exhibits a high degree of regularity.  \textcite{alessandretti2020scales} further show that individuals spend approximately 40\% of their time within a 3 km-wide area. Building on these, \textcite{ivann_2025_16913800} uses residents’ mobility traces to construct neighbourhoods by examining the cumulative overlap of users' activity spaces near their homes as a hierarchical structure based on the similarity of visited locations. The resulting clusters are compared to a POI-derived clustering relying on proximity to diversity hotspots in the city.

\subsection{Liveability assessment}
Although there is no consensus on its definition, liveability generally refers to the quality of living conditions in a given area and is affected by multiple factors \parencite{liveability_definition}. For instance, urban service diversity and proximity \parencite{moreno2021introducing}, the availability of and accessibility to urban green spaces \parencite{niemelaUsingEcosystemServices2010, tianAssessingLandscapeEcological2014}, the promotion of walkable environments \parencite{shamsuddinWalkableEnvironmentIncreasing2012}, and environmental quality \parencite{norouzian-malekiDevelopingTestingFramework2015} significantly influence urban liveability. Conversely, dispersed retail establishments \parencite{rotem-mindaliRetailFragmentationVs2012} and urban insecurity \parencite{leby2010liveability} can negatively impact liveability. 

Extensive research has focused on large-scale or city-wide liveability assessments. To this aim, The Global Liveability Index has been developed to evaluate global cities by considering diverse factors that collectively influence their overall liveability \parencite{economistintelligenceunitGlobalLiveabilityIndex2023}. While valuable for benchmarking cities across various domains, such assessments often overlook subtle variations at the neighbourhood level. 

At a more granular-level, \textcite{higgsUrbanLiveabilityIndex2019} for example, developed the Urban Liveability Index by integrating policy-related health indicators for residential addresses in Melbourne, showing its positive correlation with active travel behaviours and effectiveness in evaluating policy impacts. However, assessing solely at residential locations may misrepresent neighbourhood liveability profiles, as neighbourhoods are better conceptualised through the actual activity spaces in which daily needs are met rather than fixed home locations. \textcite{norouzian-malekiDevelopingTestingFramework2015} developed a neighbourhood liveability assessment framework by identifying key attributes that influence liveability and tailoring their weights to Estonian and Iranian neighbourhood contexts using the Delphi method. While this offers a useful conceptual basis, it does not deliver a tangible assessment in practice and relies on expert opinion to determine weights, which may constrain replicability and transferability.

The preceding literature identifies a wide range of factors contributing to liveability and its assessment. Liveability is vaguely defined and often shaped by the context of the study, such as urban health, transport access, and social segregation. Each factor reflects a distinct dimension, and no single dimension can fully capture the concept. In this study, we view liveability as an urban attractor, that is, as a feature of neighbourhoods that draws people into everyday activities and fosters active engagement. From this perspective, it is more appropriate to prioritise the functional aspects of the built environment. These aspects are spatially measurable and comparable, allowing for a clearer evaluation of neighbourhood liveability. Furthermore, the scarcity of granular-level liveability assessments underscores the need for such evaluations across diverse urban contexts. Without fine-scale assessments, important intra-urban disparities risk being overlooked, leading to one-size-fits-all policies that fail to address the unique needs and deficits of individual neighbourhoods. A detailed liveability assessment can support place-based interventions and guide more responsive neighbourhood-scale urban planning.

\section{Data}
\textbf{OpenStreetMap (OSM)}: POIs are obtained from OSM, and are used among other things to define neighbourhoods following \textcite{ivann_2025_16913800}. The road network is also obtained from OSM, and is used to compute proximity.

\begin{itemize}
\item{POIs}: In OSM, POIs are defined through key–value pairs, where the key specifies a category and the value indicates the specific feature. For this study, we focused on three keys—amenity, shop, and tourism—because they capture the majority of POIs relevant to the functional aspects of daily urban life, including services, retail, and leisure. From these keys, we manually inspected POI types by excluding disused or permanently closed POIs,    extracting 597 unique POI types. These POI types were subsequently grouped into five urban service categories-commerece, education, entertainment, healthcare and living-, following Moreno’s urban function classification \parencite{moreno2021introducing}.

\item{Road network}: OSMnx is used to obtain the road network.

\end{itemize}

\textbf{Mobility data:} GDPR compliant data provided by \textcite{locomizer}, ranging from February 1 - 28, 2023. It includes three key attributes: a unique hexagon identifier, a footfall score, and a movement modality, which is categorised as pedestrian, non-pedestrian, or all (the sum of both). Each hexagon corresponds to the Uber H3 level 10 grid (approx. 76 $m$ hexagon side length). The footfall score indicates the probability of encountering a user in a specific area during a designated time period, computed daily for each hexagon. A higher footfall score reflects greater human activity within a hexagon, whereas lower scores indicate less activity, enabling relative comparisons of movement intensity across areas and time. The footfall scores are initially normalised by the data provider by dividing each score by the maximum value in the dataset and multiplying by 100. The normalised footfall scores were then averaged over the 28-day period. While the dataset includes three movement modalities, this research focuses on pedestrian modality to investigate how neighbourhood liveability relates to active travel behaviour. Lastly, population data was derived at the Lower-layer Super Output Area (LSOA) level. LSOAs typically contain between 400 and 1,200 households and a resident population ranging from 1,000 to 3,000 people.

\section{Methodology}
\subsection{Liveability Index Development}
The Liveability Index was developed following the OECD’s 10-step technical guidelines \parencite{joint2008handbook}, adapted to the context of London and the research objectives. This study aims to provide a comprehensive neighbourhood liveability assessment, incorporating the neighbourhood construction process into the index framework. While the OECD guidelines provide a functional foundation, modifications were made to align with the research goals. In this study, good liveability is understood as neighbourhoods that offer a diverse mix of services, ensure high proximity to these opportunities, and are supported by sufficient population density to sustain them. 

Grounded in existing research and aligned with the functional aspects of daily urban life, the selected liveability dimensions incorporate the principles of the 15-minute city concept \parencite{moreno2021introducing}. Accordingly, the Liveability Index consists of three core domains and eleven indicators, each of which is measured at the neighbourhood level:

\begin{enumerate}
    
    \item \textbf{Diversity}: Captured using Shannon's entropy. POI data is grouped by service category at neighbourhood level. For each neighbourhood, POI counts and their probabilities within categories are computed to derive entropy. The diversity of living services, for example, is given by following Equation \ref{eq_entropy}:

\begin{equation}
D_{lk} = -\sum_{j}^{N_l} P_{kj} \ln{P_{kj}}
\label{eq_entropy}
\end{equation}

where $D_{lk}$ represents living service diversity in neighbourhood \textit{k}, $P_{kj}$ is the probability of amenity type \textit{j} within the living service category for neighbourhood \textit{k}. $N_{l}$ is the number of living-related amenities. Other diversity indicators follow the same approach.

\item \textbf{Proximity}: This is measured using a hybrid approach of average Euclidean and network distance. For services present within a neighbourhood, the average Euclidean distance between the neighbourhood’s centroid and each service’s centre of mass was used, reflecting the overall distribution pattern of amenities and simplifying distance calculations. For absent services, the network distance between the neighbourhood’s centroid and the nearest centre of mass in neighbouring areas was applied, accounting for residents' potential need to travel for unavailable amenities.

\item \textbf{Population Density}: LSOA-level population data was projected to the neighbourhood-level using a weighted sum approach, and then population density per square kilometre was calculated.
\end{enumerate}

\begin{table*}[ht]
\centering  
\small
\caption{Summary statistics of the eleven indicators within the three core domains: diversity, proximity and population density.\label{T2}}
\resizebox{\textwidth}{!}{ % Resize the table to the full width of the page
\begin{tabular}{lccccccccc}
\toprule
Indicator & Data unit & Count & Null values & Mean & Median & Std & Min & Max \\
\midrule
Diversity of Commerce & nats & 399 & 3 & 2.58 & 2.73 & 0.76 & 0.00 & 3.80 \\
Diversity of Education & nats & 399 & 3 & 0.66 & 0.64 & 0.45 & 0.00 & 1.88 \\
Diversity of Entertainment & nats & 399 & 2 & 1.64 & 1.73 & 0.44 & 0.00 & 2.42 \\
Diversity of Living & nats & 399 & 0 & 1.66 & 1.70 & 0.35 & 0.02 & 2.57 \\
Diversity of Healthcare & nats & 399 & 15 & 1.46 & 1.64 & 0.60 & 0.00 & 2.26 \\
Proximity to Commerce & metre & 399 & 0 & 305.79 & 212.24 & 311.42 & 6.82 & 3414.55 \\
Proximity to Education & metre & 399 & 0 & 314.71 & 231.52 & 308.05 & 10.95 & 2635.06 \\
Proximity to Entertainment & metre & 399 & 0 & 267.46 & 229.09 & 218.60 & 3.27 & 2153.25 \\
Proximity to Living & metre & 399 & 0 & 114.39 & 85.21 & 111.79 & 1.76 & 874.12 \\
Proximity to Healthcare & metre & 399 & 0 & 421.02 & 288.61 & 452.72 & 9.57 & 3070.22 \\
Population Density per km$^2$ & people/km$^2$ & 399 & 0 & 6544.81 & 5558.99 & 3948.37 & 145.46 & 19723.51 \\
\bottomrule
\end{tabular}
}
\end{table*}

Table \ref{T2} presents the summary statistics for all indicators, highlighting missing values in diversity indicators due to neighbourhoods lacking amenities in specific urban service categories. These should be distinguished from diversity scores of 0, which indicate the presence of only a single amenity type within a category.
To address this, a diversity value of -1 was assigned to neighbourhoods without any amenities in certain categories, differentiating them from those with a single amenity type (entropy = 0). This ensures all neighbourhoods are retained in the analysis and positioned at the lowest bound of the exponential distribution when applying the transformation, preserving a comprehensive liveability profile.

Once neighbourhood-level indicators were calculated, they were normalised using a rank-based exponential transformation, which rescales values to a range of 0–100. Following normalisation, indicator weights were derived from Principal Component Analysis (PCA) and applied during aggregation to construct the Liveability Index. The robustness of the index was assessed using Monte Carlo simulations and by Dirichlet distributions to account for potential uncertainties that may arise during the index development process. Finally, the Liveability Index was validated against the ‘Geographical Barriers’ sub-domain within the Barriers to Housing and Services domain of the English Indices of Deprivation 2019. Detailed steps for the index construction are provided in Appendix \ref{A2:Index}.

\subsection{Modelling}
We investigated the relationship between neighbourhood footfall patterns and the liveability profiles of neighbourhoods. In this study, the average normalised footfall score, aggregated at the neighbourhood level, was used as the dependent variable, and the liveability score was used as the indepedent variable in both the Ordinary Least Squares (OLS) and Geographically Weighted Regression (GWR) models.

\subsubsection{Ordinary Least Squares (OLS)}
Ordinary Least Squares (OLS) is a linear regression method that estimates the relationship between a dependent variable and one or more independent variables by minimising the sum of squared residuals \parencite{burton2021ols}. In this study, OLS is used to examine the global relationship between footfall and liveability scores across neighbourhoods. The model is expressed in Equation \ref{eq_ols}:

\begin{equation}
Foot_{k} = \beta_{0} + \beta_{1} LI_{k} + \epsilon
\label{eq_ols}
\end{equation}
where $Foot_{k}$ denotes the footfall score in neighbourhood \textit{k}, $LI_{k}$ is the liveability score for neighbourhood \textit{k}, $\beta_{0}$ is the intercept, $\beta_{1}$ is the coefficient for $LI_{k}$, and $\epsilon$ is the error term.

Given that the model variables are spatially referenced, we carried out Moran's I test to assess spatial autocorrelation in the model residuals. Spatial autocorrelation refers to the correlation of a variable with itself through space, capturing the degree to which nearby spatial units are similar or dissimilar \parencite{getis2009spatial}. The formula for global Moran's I is given in Equation \ref{eq_moran_i}:

\begin{equation}
I = \frac{N}{W} \cdot \frac{\sum_{i} \sum_{j} w_{ij} (x_i - \bar{x})(x_j - \bar{x})}{\sum_{i} (x_i - \bar{x})^2}
\label{eq_moran_i}
\end{equation}
where \textit{N} denotes the number of spatial units, $x_{i}$ and $x_{j}$ represent the values of the variable of interest at spatial units \textit{i} and \textit{j} respectively, $\bar{x}$ is the mean of the variable across all units, $w_{ij}$ indicates the spatial weight between locations \textit{i} and \textit{j}, and \textit{W} refers to the sum of all spatial weights in the matrix.

Global Moran's I values range from -1 to 1. A value near 1 indicates clustering (positive spatial autocorrelation), a value near -1 suggests dispersion (negative spatial autocorrelation), and a value around 0 implies spatial randomness. For this analysis, neighbours were defined using the k-nearest neighbours method. This approach was preferred over other methods, such as contiguity- or distance-based definitions for the following reasons. First, social and mobility interactions tend to cluster around a few nearby areas rather than extend evenly across all adjacent neighbourhoods. Second, the physical and artificial geography of London creates situations where areas share boundaries but are not functionally connected. We defined neighbours using \textit{k} = 4 to balance connectivity and local relevance. Smaller values risk isolating neighbourhoods, while larger ones dilute local variation by linking areas that are not functionally related. Row standardisation was then applied to generate the spatial weights matrix.

\subsubsection{Geographically Weighted Regression (GWR)}
GWR accounts for spatial non-stationarity by allowing the relationship between variables to vary across space \parencite{brunsdon1996gwr}. Spatial non-stationarity occurs when the effect of a predictor variable differs across locations — for instance, liveability may influence footfall more strongly in some neighbourhoods than in others. Instead of assuming a single global relationship across the entire study area, GWR calibrates a separate local regression model at each observation point in geographic space. The model is expressed in Equation \ref{eq_gwr}:

\begin{equation}
{Foot}_k = \beta_0(u_k, v_k) + \beta_1(u_k, v_k) \cdot {LI}_k + \varepsilon_k    
\label{eq_gwr}
\end{equation}
where ${Foot}_k$, ${LI}_k$, and $\varepsilon_k$ represent the footfall score, liveability score, and error term at location $k$, respectively. $\beta_0(u_k, v_k)$ is the spatially varying intercept, while $\beta_1(u_k, v_k)$ is the locally estimated coefficient for the liveability score at location $k$, with $(u_k, v_k)$ denoting spatial coordinates.
 
A Gaussian kernel function was used to assign spatial weights to neighboring observations, giving greater influence to those closer in space. The optimal bandwidth of 0.0075 (corresponding to approximately 3 of the 399 neighborhoods) was determined via least squares cross-validation, minimising the root mean squared prediction error of the GWR model. By capturing local variation in relationships, GWR provides a more geographically nuanced understanding of how liveability affects footfall across Greater London neighbourhoods.

The performance of both models was evaluated using the Akaike Information Criterion (AIC), its corrected form (AICc), and the adjusted $R^2$.

\section{Results}
\subsection{Liveability overview}
% liveability map
\begin{figure*}[ht]
\centering

\begin{subfigure}{0.49\textwidth}
    \centering
    \includegraphics[width=\textwidth]{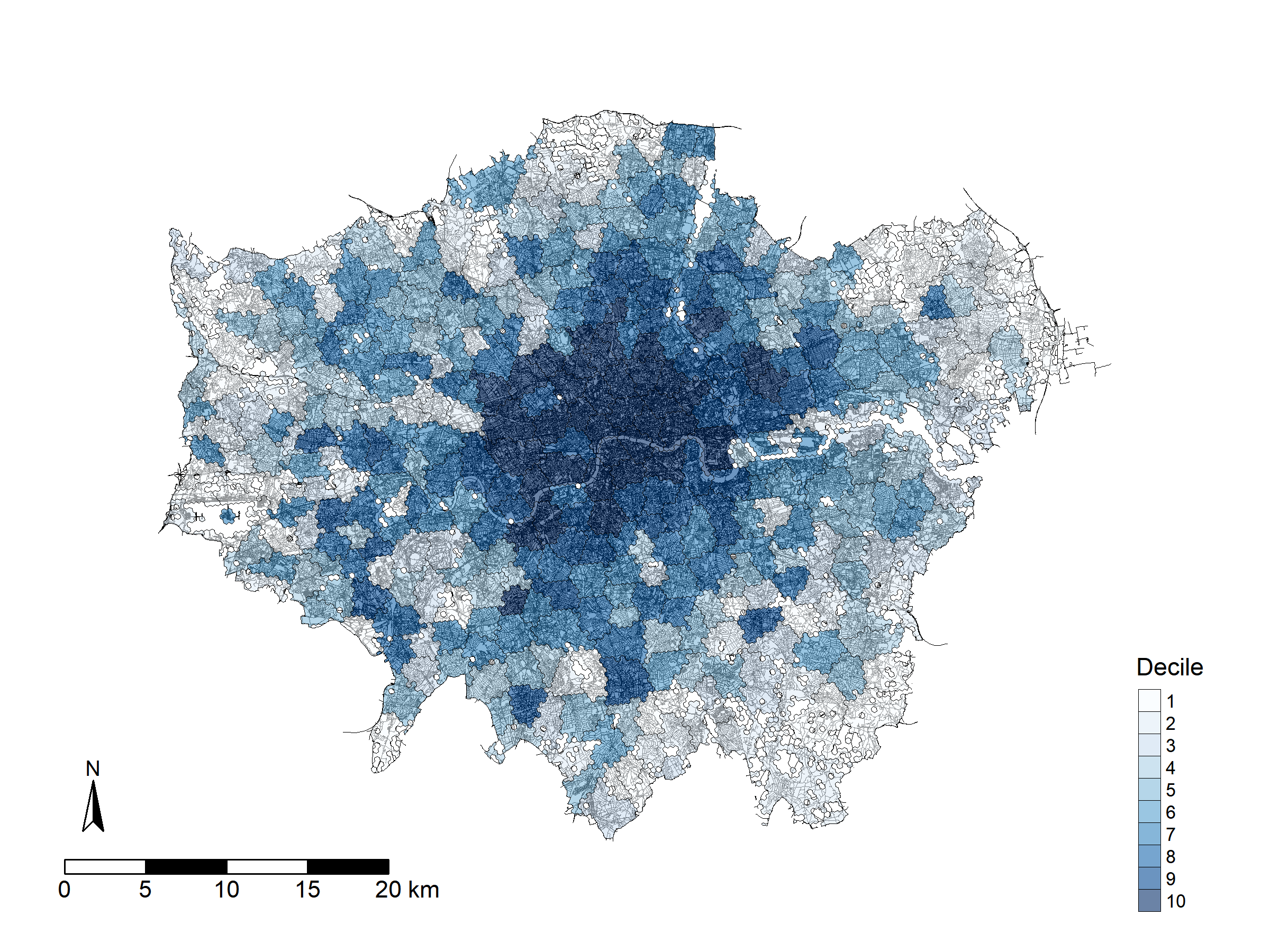}
    \caption{Overall Liveability}
    \label{overall}
\end{subfigure}
\hfill
\begin{subfigure}{0.49\textwidth}
    \centering
    \includegraphics[width=\textwidth]{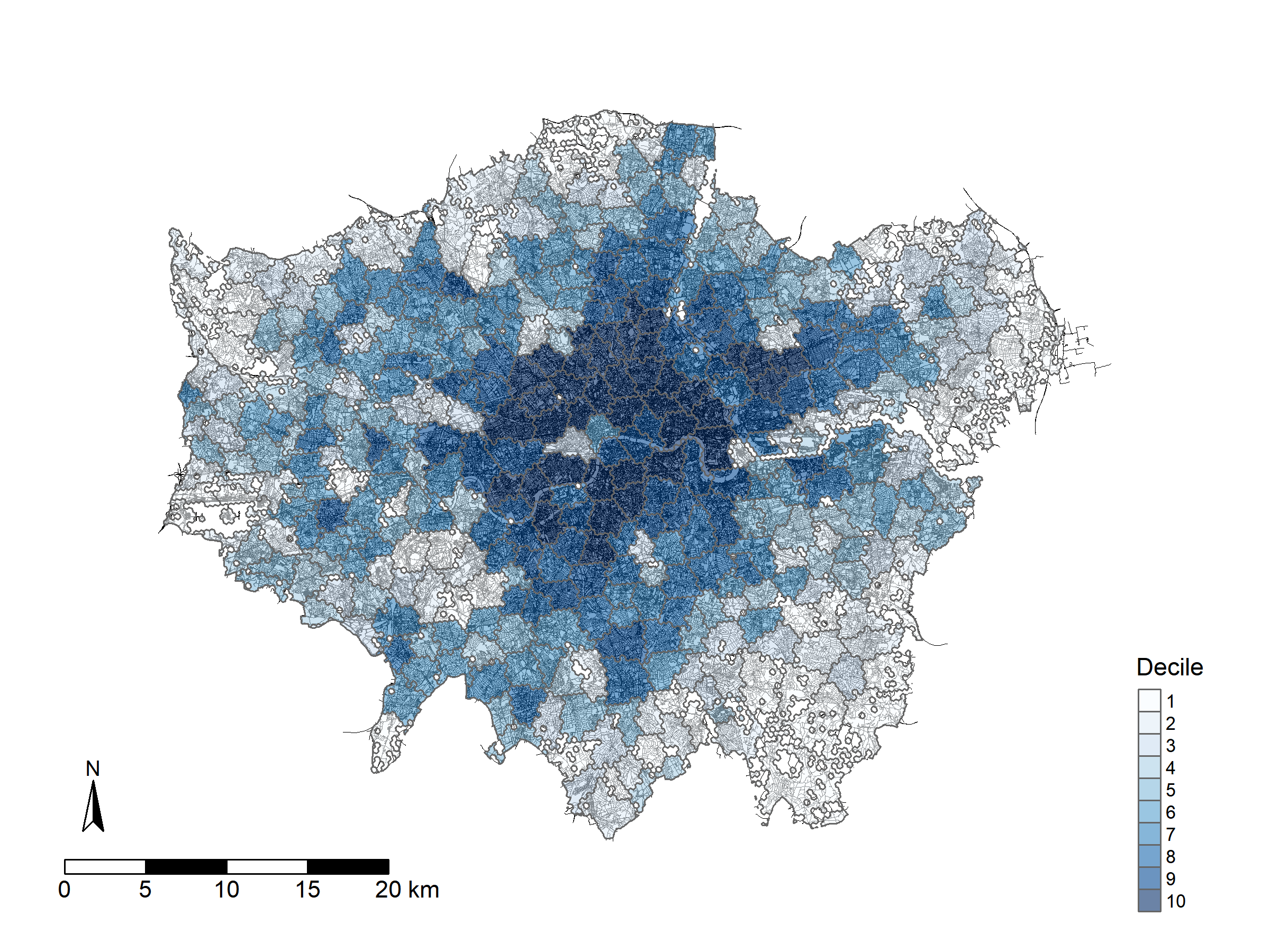}
    \caption{Population Density Domain}
    \label{density}
\end{subfigure}

\vspace{0.3cm}

\begin{subfigure}{0.49\textwidth}
    \centering
    \includegraphics[width=\textwidth]{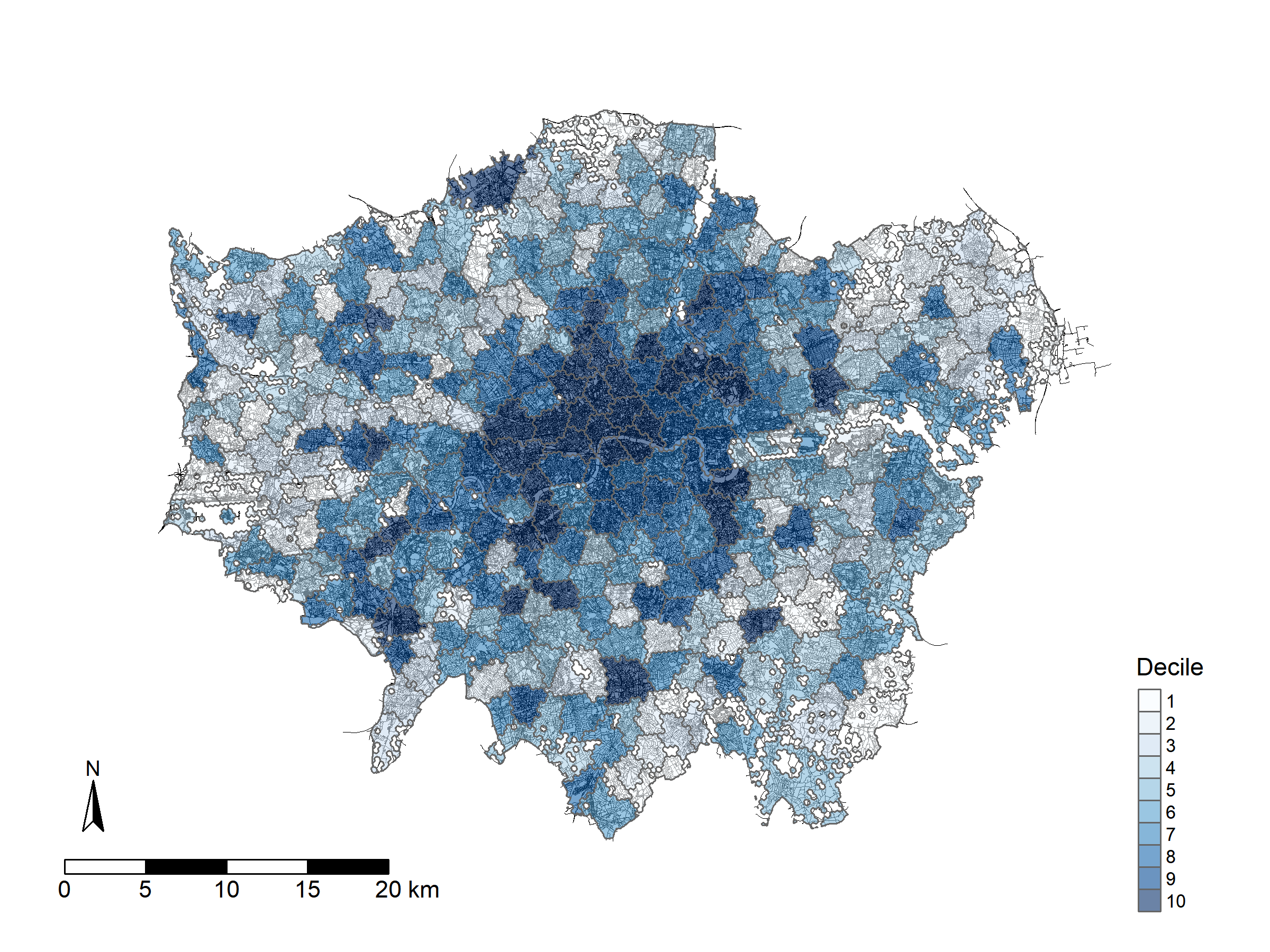}
    \caption{Diversity Domain}
    \label{diversity}
\end{subfigure}
\hfill
\begin{subfigure}{0.49\textwidth}
    \centering
    \includegraphics[width=\textwidth]{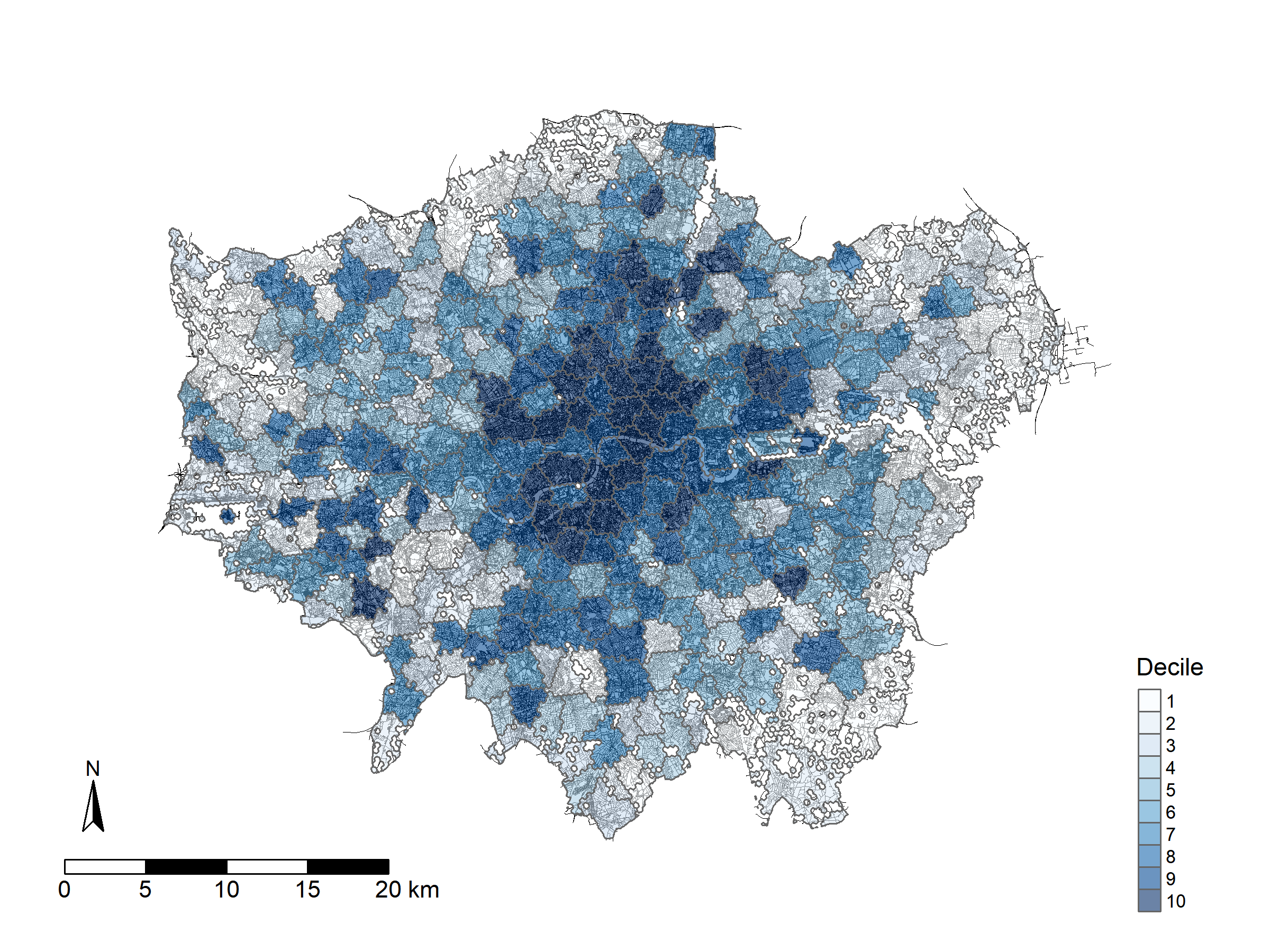}
    \caption{Proximity Domain}
    \label{proximity}
\end{subfigure}

\caption{Overall liveability and domain-specific profiles of Greater London neighbourhoods. A score of 1 indicates the lowest liveability, while 10 represents the highest.}
\label{fig:grid_4}
\end{figure*}

Figure \ref{overall} shows the liveability profiles of Greater London neighbourhoods, highlighting that Inner London generally exhibits higher liveability than Outer London. This spatial contrast can be explained by the interaction of density, diversity, and proximity, within the framework of agglomeration economies, as discussed further in the Discussion section.

Figure \ref{density}, \ref{diversity}, and \ref{proximity} illustrate the domain-specific profiles of the Liveability Index. All domain profiles exhibit broadly similar spatial patterns, with high-decile neighbourhoods clustering in Inner London and low-decile ones dominating in Outer London. This reflects the typically higher population concentration, urban service proximity, and amenity diversity found in Inner London.

However, subtle differences emerge in the diversity domain, especially in peripheral neighbourhoods. While these areas score relatively low in the population density and proximity domains, their diversity scores remain comparatively high. This pattern likely reflects the presence of a critical mass of amenities, such as regional hubs, specialised facilities, or planning strategies designed to maintain functional diversity across broader areas. As a result, the diversity domain is less dependent on local clustering of people or amenities and can remain robust even in less densely populated neighbourhoods.

\subsection{Deconstruction of Liveability Index}
Decomposing the composite index (CI) into individual indicators provides a clearer evaluation of the overall performance of a subject of interest \parencite{joint2008handbook}. This is particularly valuable as it facilitates a deeper understanding of the strengths and weaknesses within each neighbourhood. For ease of interpretation, neighbourhoods were aggregated by borough in Greater London. Due to discrepancies between neighbourhoods and administrative boundaries, each neighbourhood has been assigned to the borough with which it shares the largest intersection area.

\begin{figure*}[ht] % Use figure* for full-width spanning
\centering
\includegraphics[width=0.9\textwidth]{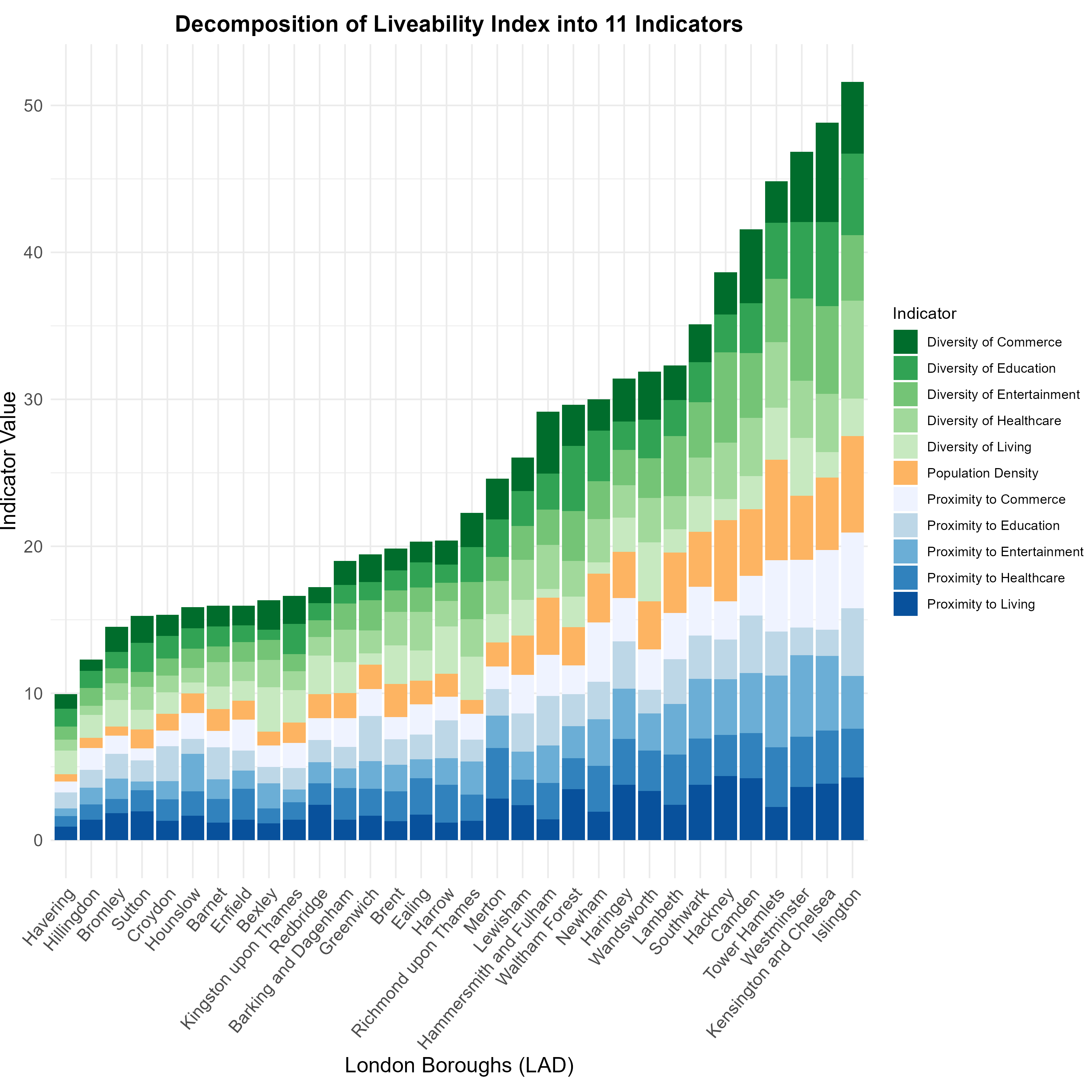} % Use full page width
\caption{Neighbourhood liveability aggregated by borough in Greater London}
\label{decomposition_li}
\end{figure*}

Figure \ref{decomposition_li} shows that the nine highest-ranked boroughs are all located in Inner London, with Islington leading as the only borough scoring above 50, followed by Kensington and Chelsea, and Westminster. In these top-ranked boroughs, the diversity domain is the most prominent contributor to the Liveability Index (LI) score. Its influence diminishes among mid-ranked boroughs, whereas in the bottom-ranked boroughs, the proximity domain appears to have a stronger impact. Population density contributes consistently around 10\% across high- and mid-ranked boroughs but becomes less significant in lower-ranked boroughs.

\begin{figure*}[ht]
    \centering
    % Top plot (a)
    \begin{subfigure}[b]{0.9\textwidth}
        \centering
        \includegraphics[width=0.9\textwidth, height=0.25\textheight]{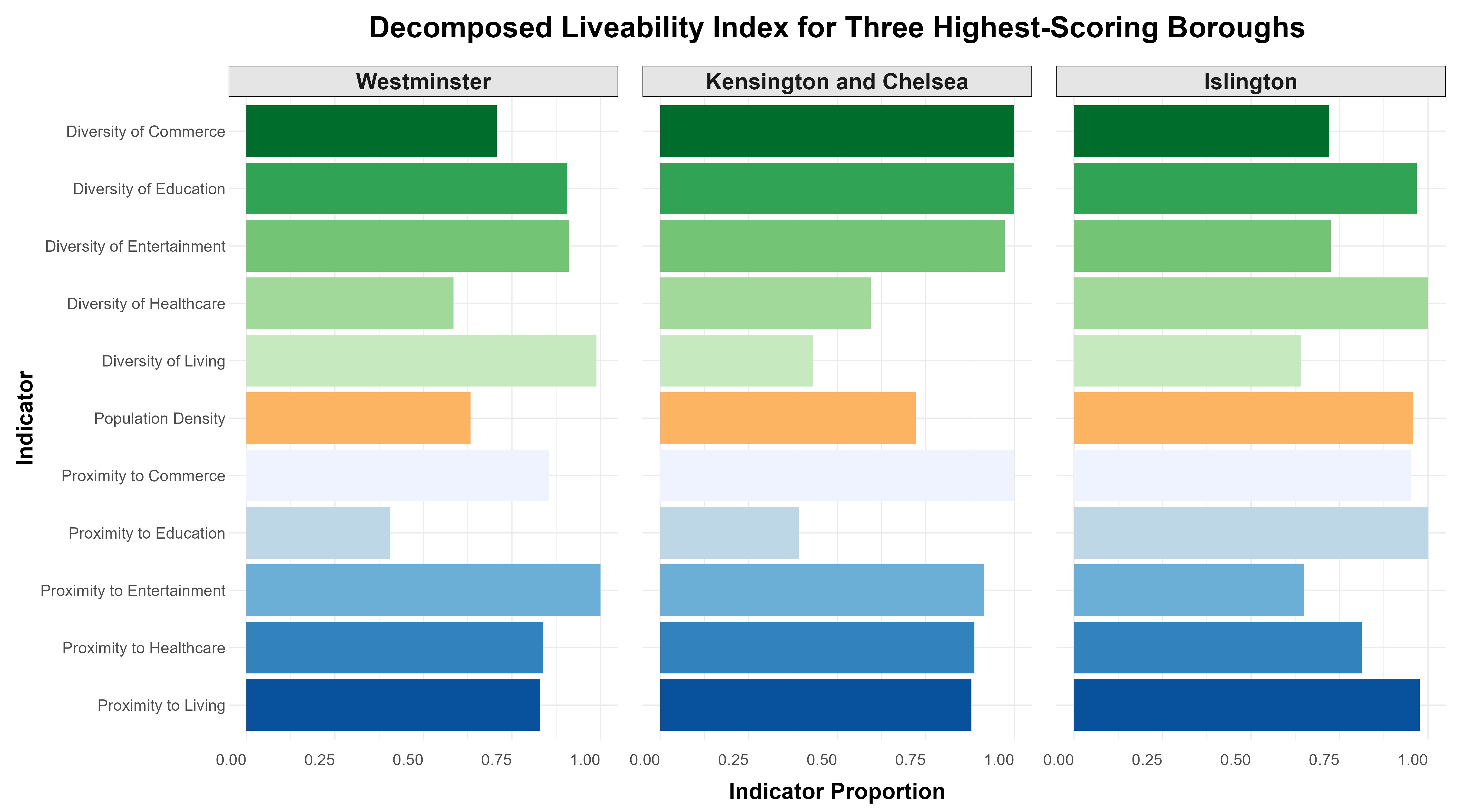}
        \caption{Three highest-scoring boroughs}
        \label{top_3}
    \end{subfigure}
    
    \vspace{0.3cm} % optional vertical space between plots

    % Bottom plot (b)
    \begin{subfigure}[b]{0.9\textwidth}
        \centering
        \includegraphics[width=0.9\textwidth, height=0.25\textheight]{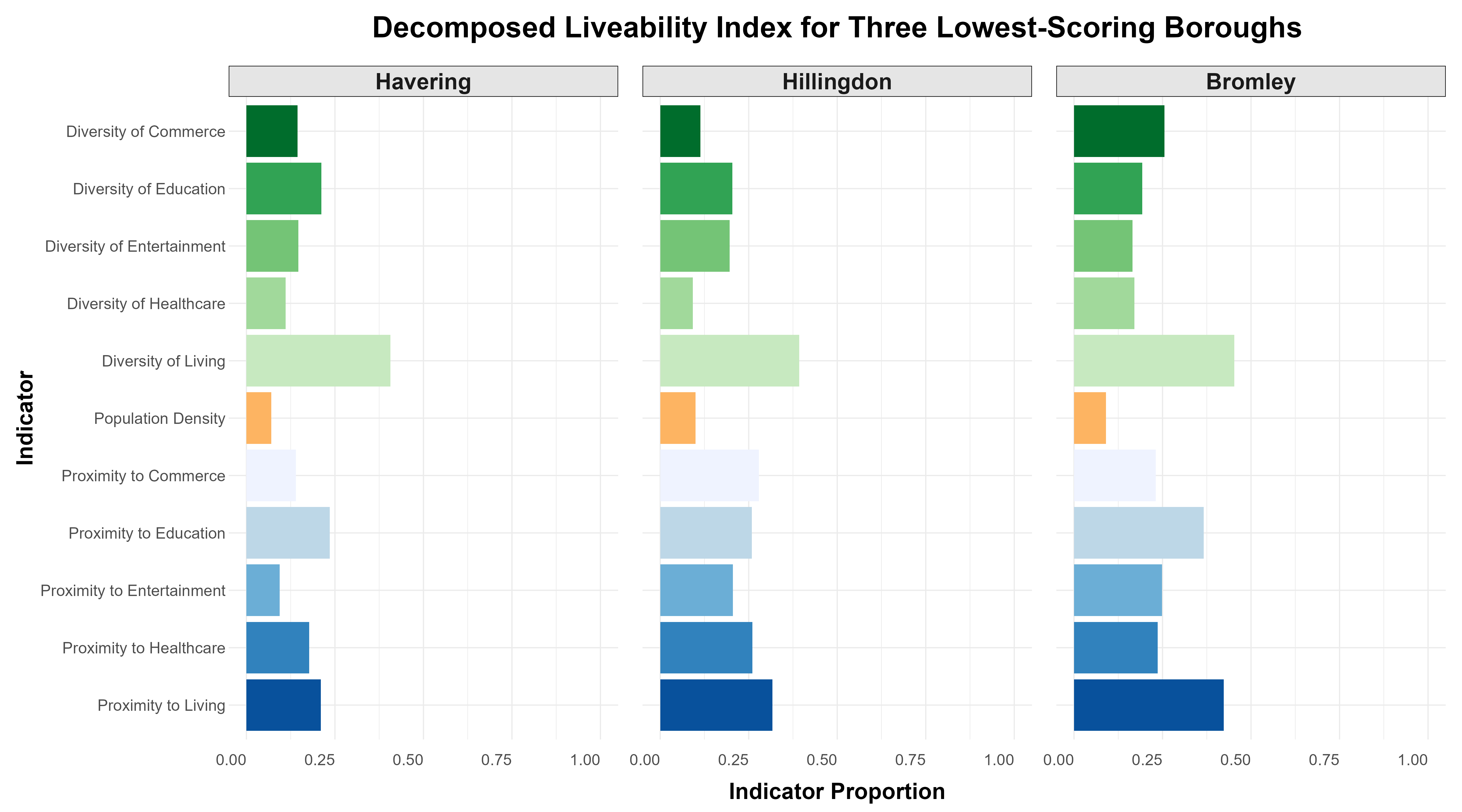}
        \caption{Three lowest-scoring boroughs}
        \label{bottom_3}
    \end{subfigure}

    % Overall figure caption
    \caption{Deconstructed indicator scores for top and bottom three boroughs. Each indicator is normalised by dividing by the maximum value of that indicator across all boroughs, allowing direct comparison of relative performance across dimensions.}
    \label{fig:top_bottom_3}
\end{figure*}

Figure \ref{fig:top_bottom_3} provides a detailed view of the individual indicator contributions for top- and bottom-ranked boroughs. Each indicator score is normalised by dividing it by the maximum score of that indicator across all boroughs. This normalisation makes it easier to compare the performance of different boroughs across multiple indicators, even if the absolute values of the scores vary significantly between indicators. In Figure \ref{bottom_3}, for instance, among the three lowest-ranked boroughs, the diversity of living stands out as the highest contributor. However, when considered collectively, the proximity indicators have a stronger overall influence on the LI score. In contrast, population density remains the least influential factor.

% decomposition of LI

% top/bottom-3 figure

\subsection{Active travel behaviour modelling}
Table \ref{ols_results} presents the results of the Ordinary Least Squares (OLS) model. Both the intercept and the liveability score are statistically significant predictors of footfall levels (\textit{p} < 0.001). Specifically, a one-unit increase in the liveability score is associated with an estimated increase of 0.0084 in footfall score. For example, neighbourhoods at the 75th percentile of liveability ($\approx 31$) are predicted to have an increase of approximately 0.21 in footfall score compared with those at the 25th percentile ($\approx6.5$). Relative to the mean footfall score of 0.496, this corresponds to an increase of about 42\%, indicating that more liveable areas tend to promote substantially higher walking activity.

The model explains a substantial portion of the variation in footfall scores, with an adjusted $R^2$ of 0.5917—indicating that approximately 59\% of the variance is accounted for by the liveability score alone. The Akaike Information Criterion (AIC) and its corrected version (AICc), both around -430, serve as indicators of model quality, where lower values signify a better fit.

% OLS results
\begin{table}[H]
\centering
\renewcommand{\arraystretch}{1.3}
\setlength{\tabcolsep}{12pt}
\caption{The results of ordinary least squares (OLS) model of neighbourhood footfall as a function of liveability. Model fit is summarised using $R^2$, adjusted $R^2$, AIC, and AICc.}
\label{ols_results}
\small
\begin{tabular}{lrrrr}
\toprule
Parameter & Coefficient & Std. Error & \textit{T-statistics} & \textit{P-value} \\
\midrule
Intercept & 0.3134 & 0.0104 & 30.27 & 2e-16 \\
Liveability Score & 0.0084 & 0.0003 & 24.04 & 2e-16 \\
$R^2$ & 0.5928 \\
Adjusted $R^2$ & 0.5917 \\
AIC & -430.8576 \\
AICc & -430.8273 \\
\bottomrule
\end{tabular}
\vspace{1pt}
\end{table}

In Table \ref{tab_moran}, Moran’s I statistic (0.3557, \textit{p} < 0.001) reveals a statistically significant positive spatial autocorrelation in the model residuals. This suggests that similar residual values—either high or low—tend to cluster geographically, rather than being randomly distributed across space. The strong positive \textit{Z-score} of 10.601 further confirms that the observed spatial pattern is highly unlikely to have occurred by chance.

% Moran's I Spatial Autocorrelation Test Table
\begin{table}[H]
\centering
\renewcommand{\arraystretch}{1.3}
\setlength{\tabcolsep}{12pt} % Adjust spacing as needed
\caption{\centering The results of Moran's I spatial autocorrelation test on OLS residuals. Global Moran’s I ranges from –1 to 1, where values near 1 indicate clustering of similar values, values near –1 indicate dispersion of dissimilar values, and values near 0 suggest a random spatial pattern.}
\label{tab_moran}
\small
\begin{tabular}{ccccc}
\toprule
Moran's I & Expectation & Variance & \textit{Z-score} & \textit{P-value} \\
\midrule
0.3558 & -0.0025 & 0.0011 & 10.602 & 2.2e-16 \\
\bottomrule
\end{tabular}
\vspace{1pt}
\end{table}

% GWR results
\begin{table}[ht]
\centering
\renewcommand{\arraystretch}{1.3}
\setlength{\tabcolsep}{12pt}
\caption{The results of geographically weighted regression (GWR) model of neighbourhood footfall as a function of liveability. Reported are the minimum, first quartile (Q1), median, third quartile (Q3), and maximum values of local parameter estimates across neighbourhoods, together with the global coefficient which is identical to that of the OLS model. Model fit is summarised using $R^2$, adjusted $R^2$, AIC, and AICc. }
\label{tab:gwr_coefficients}
\small
\begin{tabular}{lrrrrrr}
\toprule
Parameter & Minimum & Q1 & Median & Q3 & Maximum & Global\\
\midrule
Intercept & 0.0456 & 0.2593 & 0.3273 & 0.3803 & 1.0576 & 0.3134\\
Liveability Score & -0.0005 & 0.0067 & 0.0086 & 0.0117 & 0.0249 & 0.0084\\
$R^2$ & 0.8341 \\
Adjusted $R^2$ & 0.7815 \\
AIC & -698.2946 \\
AICc & -535.2158 \\
\bottomrule
\end{tabular}
\vspace{1pt}
\end{table}

\begin{figure}[ht]
\centering

\begin{subfigure}{0.49\textwidth}
    \centering
    \includegraphics[width=\textwidth]{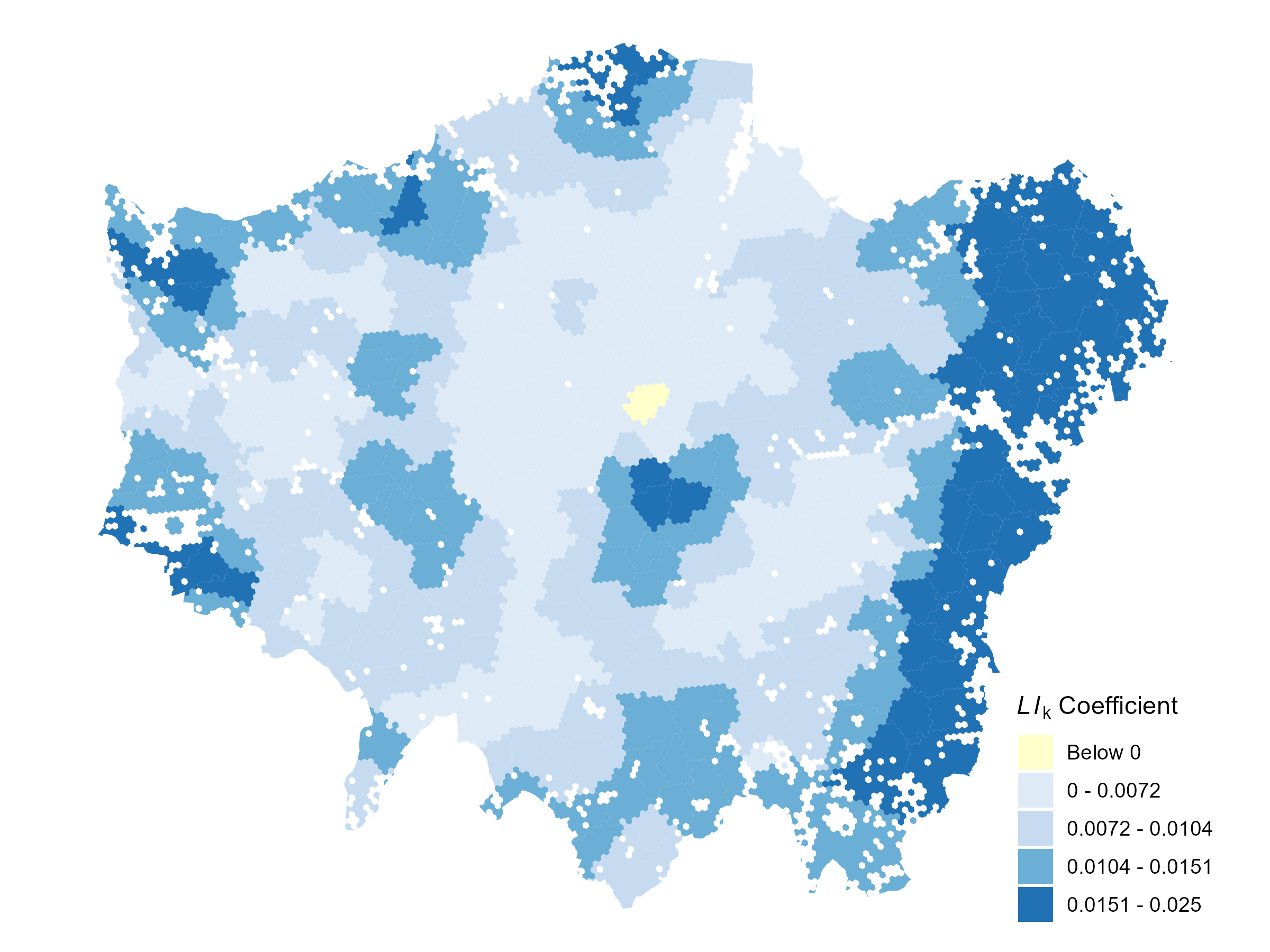}
    \caption{Liveability Score Coefficients}
    \label{gwr_coef}
\end{subfigure}
\hfill
\begin{subfigure}{0.49\textwidth}
    \centering
    \includegraphics[width=\textwidth]{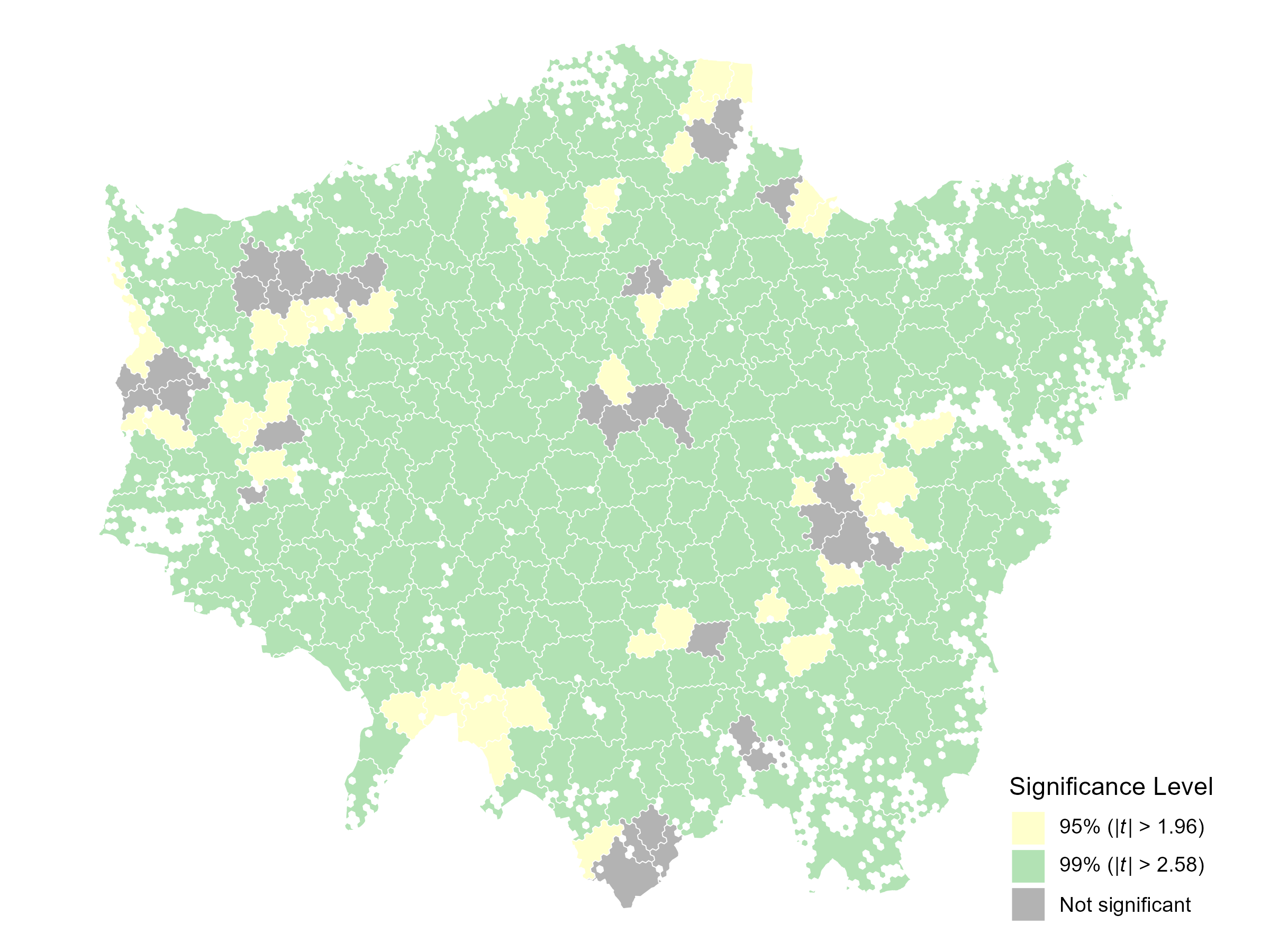}
    \caption{Pseudo \textit{t}-values}
    \label{gwr_coef_sig}
\end{subfigure}

\caption{Spatial distribution of liveability score coefficients and their pseudo \textit{t}-values}
\label{fig:gwr}
\end{figure}

The results of the Geographically Weighted Regression (GWR) model are summarised in Table \ref{tab:gwr_coefficients}. A key feature of the GWR output is the spatial variability in the liveability score’s influence on footfall. The coefficient for liveability ranges from -0.0005 to 0.0249. In Figure \ref{gwr_coef}, the spatial distribution of the liveability score coefficients demonstrates considerable variation across Greater London. Most neighbourhoods show a positive association between liveability and footfall, suggesting that more liveable environments generally support higher levels of footfall activity. Interestingly, the strongest positive coefficients are observed in peripheral areas, where overall liveability profiles tend to be lower. This may imply that in less liveable areas, even marginal improvements in liveability could lead to relatively larger gains in footfall.

Figure \ref{gwr_coef_sig} presents the pseudo \textit{t}-values, calculated by dividing each local coefficient by its standard error, to indicate statistical significance. Pseudo \textit{t}-values greater than 1.96 are typically considered statistically significant at the 95\% confidence level. The map shows that while not all local coefficients are significant, a substantial number of neighbourhoods exhibit statistically robust relationships between liveability and footfall, reinforcing the relevance of local variation captured by the GWR model.

% GWR figure

The model achieves a notably high adjusted $R^2$ of 0.7815, indicating that approximately 78\% of the variation in neighbourhood footfall scores is explained by local variations in liveability. Furthermore, both AIC and AICc values, -698.2946 and -535.2158 respectively, are substantially lower than those of the OLS model, suggesting superior model fit. The GWR residuals were assessed using Moran’s I test and residual-versus-fitted plots, both of which indicated no signs of spatial autocorrelation or systematic bias, suggesting that the model is well-specified and does not suffer from overfitting (see Appendix \ref{A3:MoransI}).

While both OLS and GWR models indicate a generally positive relationship between liveability and footfall, the GWR model reveals that this relationship is context-dependent and shaped by local dynamics. It is important to note, however, that these findings reflect statistical associations and do not establish causality.

\section{Discussion}
\subsection{Liveability profiles of Greater London neighbourhoods}
The distinct spatial patterns of liveability in Greater London neighbourhoods arise from the interplay of three core domains: diversity, proximity, and population density. Inner London’s high population creates a heightened demand for urban services, spurring the development of diverse amenities. Consequently, residents in these neighbourhoods have convenient access and proximity to essential urban services. These patterns align with the framework of agglomeration economies, which posits that population and economic clustering drive urban growth and the development of amenities \parencite{glaeser1992growth}. Ultimately, the intricate interplay among the domains culminates in the elevated liveability observed in Inner London neighbourhoods. It is interesting to note that in peripheral neighbourhoods, the diversity domain remains relatively resilient, supported by strategically important amenities that serve broad catchment areas. Facilities such as regional hospitals, major shopping centres, and leisure complexes act as focal points for surrounding communities, providing essential services and recreational opportunities even in areas of lower population density. Kingston, located in the south-west of London, which, as shown in \ref{fig:grid_4}.c, has a relatively high diversity score, provides a clear example: Kingston Hospital delivers specialised health services to a wide catchment, while retail complexes like the Bentall Centre and Eden Walk concentrate shops and services, supporting a varied mix of daily needs. Higher-education campuses and leisure facilities further expand the functional mix, and well-connected transport links extend the reach of these amenities across neighbouring areas. Together, these regional hubs help sustain this aspect of liveability, highlighting the importance of strategic investment in key facilities to maintain robust diversity across outer London neighbourhoods.

Breaking down liveability scores into individual indicators provides deeper insights into each borough’s strengths and weaknesses, offering policymakers data-driven guidance. The three boroughs with the lowest liveability scores–Havering, Hillingdon, and Bromley-could particularly benefit from integrating these findings into local planning. Taking Havering as an example, Figure \ref{bottom_3} highlights its key challenges in proximity to entertainment and commerce, and diversity of healthcare, as reflected in its local planning reports. Poor north-south transport connectivity \parencite{havering2021} and the geographic alignment of district centres along the same axis limit access to commercial and recreational facilities \parencite{havering2018}. With regards to healthcare, Havering experiences significant spatial health inequalities driven by deprivation in the north and south \parencite{havering2021}, alongside a pressing need for larger medical facilities, with an estimated requirement of approximately 18,000 m$^2$ compared to the current 7,179 m$^2$, and an average of 2,690 patients per GP, far exceeding the London average of 2,100 and the national average of 2,000 \parencite{erm2018}.

These instances highlight the Liveability Index’s potential as both a diagnostic tool for evidence-based policymaking and a means of monitoring policy implementation. For example, ‘Policy 16: Social Infrastructure’ aims to improve residents’ access to essential services through public transport and active travel modes while fostering new social infrastructure via mixed-use development \parencite{havering2021}. Tracking changes in proximity and diversity indicator scores can help evaluate policy effectiveness. Furthermore, as resident participation becomes increasingly vital in shaping neighbourhoods, the index’s easy-to-understand feature can enhance public understanding of local challenges and encourage greater community engagement in urban planning.

\subsection{Correlation between liveability and footfall patterns}
Footfall Modelling offers valuable insights into how neighbourhood characteristics influence active travel behaviour. The findings underscore the potential influence of liveability of urban settings on individuals’ active travel choice. This aligns with existing research that highlights that urban designs prioritising people’s needs and accessibility foster increased physical activity \parencite{hooperBuildingBlocksLiveable2015}, and the positive correlation between liveability and active travel frequencies \parencite{higgsUrbanLiveabilityIndex2019}, further reinforcing the study's outcomes.

The relatively strong performance of the OLS model, alongside the notably high local explanatory power of the GWR model, underscores the effectiveness of the three core liveability domains—diversity, proximity, and density—in capturing spatial variations in footfall patterns. These dimensions appear sufficient for modelling active travel behaviour, suggesting that they encapsulate the key spatial and functional qualities that support walkability and accessibility in urban settings.

Importantly, the spatial heterogeneity in liveability coefficients revealed by the GWR model reinforces the non-stationary nature of the relationship between liveability and footfall. That is, the strength and direction of the association vary across space. In some neighbourhoods, high liveability scores may not necessarily translate into high footfall, possibly due to contextual factors such as demographic characteristics (e.g., older populations) or the quality of public realm infrastructure. Conversely, in areas where walkability and access to amenities are well integrated, the positive impact of liveability on footfall is more pronounced.

Nevertheless, it is crucial to recognise the multifaceted nature of active travel behaviour. While neighbourhood liveability plays a significant role, the decision to opt for active travel modes is influenced by a complex interplay of factors, these include topological and climatic conditions \parencite{wismadi2025sustainable}, the quality of active travel infrastructure \parencite{timmons2024active}, policy interventions \parencite{winters2017policies}, and socio-economic circumstances \parencite{younkin2024influence}.

\subsection{Limitations}
To help interpretation, this research aggregated neighbourhoods at the borough level. However, this approach may inadvertently misrepresent liveability characteristics at both neighbourhood and borough scales. Neighbourhoods intersecting multiple boroughs were allocated to the borough with the largest overlap. Since administrative boundaries do not always align with neighbourhood divisions, this aggregation method might have led to an overestimation or underestimation of certain boroughs’ liveability profiles.

Furthermore, a diversity value of -1 was imputed for neighbourhoods lacking specific urban amenities to ensure a comprehensive representation of urban service diversity. While this imputation maintains the data distribution by placing neighbourhoods with missing amenities at the lowest bound, it is important to note that the minimum entropy value remains 0. Future research could explore alternative approaches for measuring diversity more accurately.

While GWR effectively models spatial variation in the relationship between liveability and footfall, it assumes this relationship is temporally static. However, footfall patterns are often shaped by time-dependent factors such as the day of the week, seasonal variation, or time of day. In this study, the footfall data was collected in February, which may not capture behavioural shifts that occur in other seasons—such as increased outdoor activity in warmer months. As such, the GWR model may not fully reflect the temporal fluctuations in travel behaviour. Future research could consider using more advanced modelling approaches that capture both spatial and temporal dynamics, such as Geographically and Temporally Weighted Regression (GTWR) or spatio-temporal Bayesian models.

\section{Conclusion}
The growing importance of cities in addressing urban challenges, alongside the shift toward human-centric urban planning, has brought the concept of the liveable neighbourhood to the forefront. Yet, varying definitions of neighbourhoods and the multifaceted nature of liveability continue to present challenges for urban planners seeking to define and assess neighbourhood liveability.

In response, this research adapts the 15-minute city paradigm to the London context and, in alignment with local policies, identifies three core domains—diversity, proximity, and population density—as fundamental components of neighbourhood liveability. These dimensions, grounded in the functional aspects of daily urban life, suffice to capture the urban attraction elements, validated by footfall data.
 
The Liveability Index developed in this study enables a fine-grained, neighbourhood-level assessment, revealing localised liveability strengths and gaps that are often obscured in broader-scale analyses. In particular, our neighbourhood-scale focus allows to concretise neighbourhood liveability as a set of social attractions that shape and sustain daily activity. This granular perspective offers urban planners deeper insights into neighbourhood dynamics and supports the implementation of targeted, place-based interventions that promote diverse, walkable, and liveable communities. Furthermore, the positive association between neighbourhood liveability and footfall activity underscores the tangible impact of neighbourhood profiles in shaping mobility patterns.

This research provides a concrete demonstration of how the long-standing vision of diverse and walkable neighbourhoods, articulated by Jane Jacobs and more recently by the 15-minute city paradigm, can be put into practice. Through a transparent index modelling process and the use of open data, the Liveability Index is shown to be adaptable and scalable to a wide range of urban contexts. By providing a holistic yet spatially grounded perspective on neighbourhood liveability, this study contributes to the development of urban spaces that enhance residents’ well-being and quality of life.

% references
\printbibliography
\newpage
\section*{Appendix}
\setcounter{subsection}{0} % reset numbering
\renewcommand{\thesubsection}{A\arabic{subsection}}
\renewcommand{\thefigure}{A\arabic{figure}}
\setcounter{figure}{0}
\renewcommand{\thetable}{A\arabic{table}}
\setcounter{table}{0}

\subsection{Data and code availability}\label{A1:onlineMaterial}
POIs dataset is available from OpenStreetMap. Population data at LSOA level is publicly available. We do not have permission to share the mobility dataset obtained from \textcite{locomizer}. The main code used to replicate the results is available at: \url{https://github.com/phily5051/liveability-index.git}

%\newpage

\subsection{Liveability Index development}\label{A2:Index}
\subsubsection{Theoretical framework}
The Liveability Index builds on the principles of the 15-minute city concept. \textcite{moreno2021introducing} identified four key dimensions—diversity, proximity, population density, and digitalisation—designed to ensure access to six essential urban services: commerce, education, entertainment, living, and healthcare. The aim of the Liveability Index is to capture liveability as a measure of urban attractors using a minimal set of variables that are commonly available across different urban contexts. To this end, the ‘working’ service is excluded, given the rise of teleworking in the post-pandemic era \parencite{vargas2022rise}, and the digitalisation domain, while central to the 15-minute city framework, is not incorporated into the index. Policy frameworks such as the London Plan 2021 and the National Planning Policy Framework (See Table \ref{policy_quote}) further justify the choice of three domains by demonstrating their alignment with established urban planning priorities. While these documents are rooted in the London context, they reflect policy directions that are also relevant across other European countries. Table \ref{li_domains_services_benefits} incorporates the quotations from Table \ref{policy_quote}, organising them into 3 domains with 11 urban services and their associated benefits. Each service accounts for distinctive dimensions within its domain, facilitating a comprehensive presentation of liveability profiles. 

\begin{table*}[htbp]
\centering
\caption{A list of the 15-minute city concept related quotes from the policy frameworks}
\label{policy_quote} % Add label here
\renewcommand{\arraystretch}{1.3} % Increase row spacing
\resizebox{\textwidth}{!}{  % This resizes the entire table to the width of the page
\begin{tabular}{l p{12cm} p{1cm} p{2.5cm}} % Adjust column width as needed
\toprule
\textbf{Domain} & \textbf{Policy Reference (quoted)} & \textbf{Page} & \textbf{Policy Document} \\
\midrule
\textbf{Diversity} & ‘Encouraging a mix of land uses, and co-locating different uses' & p.15 & London Plan \\
 & ‘High-density, mixed-use places support the clustering effect of businesses' & p.15 & London Plan \\
 & ‘Creating vibrant and active places and ensuring a compact and well-functioning city' & p.16 & London Plan \\
 & ‘Identify and allocate a range of sites to deliver housing locally' & p.22 & London Plan \\
 & ‘Promote and support London's rich heritage and cultural assets' & p.24 & London Plan \\
 & ‘Ensure the public realm is well-designed, safe, accessible, inclusive' & p.134 & London Plan \\
 & ‘Maximise the extended or multiple use of educational facilities' & p.223 & London Plan \\
 & ‘Sufficient supply of good quality sports and recreation facilities' & p.229 & London Plan \\
 & ‘A successful, competitive and diverse retail sector, which promotes sustainable access to goods and services for all Londoners' & p.267 & London Plan \\
 & ‘Support strong, vibrant and healthy communities' & p.5 & NPPF \\
 & ‘Enable and support healthy lifestyles … through the provision of safe and accessible green infrastructure, sports facilities, local shops, access to healthier food' & p.27 & NPPF \\
 & ‘To provide the social, recreational and cultural facilities and services the community needs' & p.27 & NPPF \\
\midrule
\textbf{Proximity} & ‘Mayor's target for 80 percent of all journeys to be made by walking, cycling and public transport' & p.15 & London Plan \\
 & ‘Support active travel' & p.114 & London Plan \\
 & ‘Social infrastructure should be easily accessible by walking, cycling and public transport' & p.218 & London Plan \\
 & ‘Improving street environments to make walking and cycling safe and more attractive' & p.402 & London Plan \\
 & ‘Reduce car dominance' & p.402 & London Plan \\
 & ‘Use of attractive, well-designed, clear and legible pedestrian and cycle routes' & p.27 & NPPF \\
\midrule
\textbf{Digitalisation} & ‘Fast, reliable digital connectivity is essential' & p.362 & London Plan \\
 & ‘Support the expansion of electronic communications networks' & p.33 & NPPF \\
\midrule
\textbf{Population Density} & ‘Density measures are related to the residential population will be relevant for infrastructure provision' & p.116 & London Plan \\
\bottomrule
\end{tabular}
}  % End the resizebox
\end{table*}

%table1
\begin{table*}[h]
\centering
\caption{Liveability Index domains, services and benefits}
\label{li_domains_services_benefits} % Add label here
\resizebox{\textwidth}{!}{  % This resizes the entire table to the width of the page
\begin{tabular}{l p{5cm} p{10cm}} % Adjust column width as needed
\toprule
\textbf{Domain} & \textbf{Service} & \textbf{Benefit} \\
\midrule
\textbf{Diversity} & Diversity of Commerce & Enhanced variety of commerce-related amenities and activities \\
 & Diversity of Education & Enhanced variety of education-related amenities and activities \\
 & Diversity of Entertainment & Enhanced variety of entertainment-related amenities and activities \\
 & Diversity of Living & Enhanced variety of living-related amenities and activities \\
 & Diversity of Healthcare & Enhanced variety of healthcare-related amenities and activities \\
\midrule
\textbf{Proximity} & Proximity to Commerce & Improved proximity to commerce-related amenities and activities \\
 & Proximity to Education & Improved proximity to education-related amenities and activities \\
 & Proximity to Entertainment & Improved proximity to entertainment-related amenities and activities \\
 & Proximity to Living & Improved proximity to living-related amenities and activities \\
 & Proximity to Healthcare & Improved proximity to healthcare-related amenities and activities \\
\midrule
\textbf{Population Density} & Population density per ${km}^2$ & Optimal population density to sustain urban functions and resource consumption \\
\bottomrule
\end{tabular}
}  % End the resizebox
\end{table*}

\subsubsection{Data Selection and Imputation}
This section details how eleven variables, outlined in Table \ref{li_domains_services_benefits}, were measured at the neighbourhood level. Both the diversity and proximity domains contain five metrics relating to essential urban services, while population density domain captures population density per square kilometre. All variables were computed for every neighbourhood.

The diversity domain was measured using Shannon’s entropy, which captures the distribution of amenities within each urban service category. Amenity data were obtained from OpenStreetMap, covering five categories of urban services: commerce, education, entertainment, healthcare, and living (See Table \ref{poi_grouping}). For each neighbourhood, a diversity value was calculated across these five urban service types, providing an indicator of the balance and variety of amenities. In cases where neighbourhoods lacked amenities in a given urban service category, a diversity value of –1 was assigned. This distinction enabled differentiation between neighbourhoods with no amenities in a service category and those with only a single amenity, the latter being represented by a diversity score of 0. This approach facilitates the calculation of diversity indicators for all neighbourhoods, ensuring that neighbourhoods without amenities are still represented in the liveability profiles.

\begin{table}[h]
\centering
\caption{Grouping of POIs into five essential urban service categories. Each subgroup lists representative POIs.}
\label{poi_grouping}
\begin{tabular}{p{3cm}p{11cm}}
\hline
\textbf{Urban Service} & \textbf{Amenity} \\
\hline
Living & 
Transport \& mobility services (e.g., bus stop, taxi, parking, bicycle facilities, ferry terminal), 
public \& civic services (e.g., post office, townhall, courthouse, police, fire station, housing office), 
community \& social infrastructure (e.g., community centre, retirement home, hostel, coworking space), 
religious \& cultural facilities (e.g., place of worship, monastery, church hall), 
utilities \& street infrastructure (e.g., recycling, toilets, benches, water points, charging stations). \\
\hline
Commerce & 
Retail \& consumer services (e.g., supermarkets, department stores, markets, boutiques, charity/second-hand shops), 
specialised retail (e.g., books, clothing, electronics, furniture, jewellery, antiques, cosmetics, toys, musical instruments, pet shops), 
food \& beverage retail (e.g., bakeries, frozen food, off-licence, specialty food shops), 
financial \& professional services (e.g., banks, ATMs, bureau de change, estate agents, legal services), 
vehicle-related services (e.g., fuel stations, garages, car repair, bicycle shops, car wash), 
personal \& household services (e.g., hairdressers, tailors, laundry, locksmiths, florists, tattoo studios). \\
\hline
Healthcare & 
Medical \& clinical services (e.g., hospitals, clinics, doctors, dentists, pharmacies, opticians), 
animal \& veterinary services (e.g., veterinary practices, shelters), 
wellness \& alternative medicine (e.g., massage, herbalists, meditation, therapy, nutritionists), 
fitness \& lifestyle (e.g., gyms, dojos, tanning, residential care homes). \\
\hline
Education & 
Formal education (e.g., schools, kindergartens, universities, colleges), 
informal \& specialist learning (e.g., language schools, cookery schools, art/music schools, adult education, driving schools), 
cultural \& knowledge institutions (e.g., libraries, archives, research institutes). \\
\hline
Entertainment & 
Food \& drink venues (e.g., pubs, cafes, restaurants, bars, bakeries, fast food), 
arts \& culture (e.g., museums, galleries, theatres, cinemas, cultural \& exhibition centres), 
leisure \& recreation (e.g., parks, theme parks, swimming pools, sports clubs, arenas, leisure centres), 
nightlife \& social spaces (e.g., nightclubs, casinos, hookah lounges, social clubs, private clubs), 
tourism \& attractions (e.g., zoos, aquariums, historical buildings, viewpoints, amusement parks). \\
\hline
\end{tabular}
\end{table}

For proximity domain indicators, the centroid of each neighbourhood was used as the origin, and the centre of mass of each urban service within the neighbourhood as the destination. The average Euclidean distance between these points was calculated, capturing the typical distance from the neighbourhood centre to its amenities by service category. Using the neighbourhood centroid as the origin provides a reasonable approximation of the overall accessibility, while the centre of mass of urban services accounts for their spatial distribution within the neighbourhood. This approach effectively measures the collective proximity of amenities to the heart of each neighbourhood. For neighbourhoods lacking a particular urban service, the network distance between the centroid of the neighbourhood and the nearest centre of mass of that service in other neighbourhoods was calculated.

For population density, the population data were obtained at the Lower-layer Super Output Area (LSOA) level from the ONS 2020 census data. Since neighbourhood boundaries do not align with administrative boundaries, a direct projection of the LSOA-level data onto neighbourhoods could introduce inconsistencies in data interpretation. To address this, hexagon grids at the Uber H3 level 10 were employed as an intermediary level, providing a consistent basis for aggregation and enabling a more reliable comparisons between neighbourhoods. In projecting population data, a weighted sum approach was applied. This involves summing population values after proportionately multiplying each by its corresponding intersection ratio, thereby representing its weighted contribution. This method effectively accounts for the influence of each LSOA’s spatial coverage on a specific hexagon grid cell and aligns with the inherently additive nature of population counts. Once projected onto the hexagon grid level, population data were further aggregated to the neighbourhood level using the same weighted sum approach, culminating in the calculation of population density per square kilometre.

\subsubsection{Neighbourhood construction}
This study employs a diversity-based clustering methodology that uses network proximity to a mix of urban amenities \parencite{ivann_2025_16913800} to identify neighbourhoods. This approach relies on isodistance and local maxima, enabling the mapping of amenities to accessible, diverse locations. By linking amenities to these points, we can detect potential neighbourhood centres, facilitating their organic formation.

Existing literature provides guidance on determining an appropriate isodistance value. A significant share of neighbourhood trips occurs within a 10-minute walking radius \parencite{doi:10.1177/0739456X14550401}, while an amenity’s influence diminishes by half approximately every 62.5m and becomes minimal at around 500m \parencite{hidalgoAmenityMixUrban2020}. Building on these findings, this study constructs neighbourhoods based on the clustering-based approach outlined by \textcite{ivann_2025_16913800}.

\subsubsection{Normalisation}
The normalisation approach in this research serves two key purposes: aligning diverse indicators onto a common scale and addressing the compensability issue when aggregating them into a Composite Index (CI). A key consideration in the construction of the CI is the substitutability of indicators – whether a low value in one indicator can be counterbalanced by a high value in another \parencite{mazziottaMethodsConstructingNonCompensatory2016}. Given that the objective of this research is to comprehensively delineate neighbourhood liveability and enable comparisons of liveability scores, a non-compensatory aggregation approach is more suitable.

The Ministry of Housing, Communities and Local Government adopted a non-compensatory approach in formulating the Index of Multiple Deprivation (IMD), using an exponential transformation within a weighted cumulative model to minimise cancellation effects  \parencite{mclennanEnglishIndicesDeprivation2019}. This method ranks areas by indicator scores, rescaling them between 0 and 1, where the lowest indicator score corresponds to \textit{R} = 1/\textit{N} and the highest to \textit{R} = 1 (\textit{N} = 399 in this case study). Rankings are then transformed exponentially using the following Equation \ref{secondequation}:

\begin{equation}
X\ =\ -23\ \ln{(1\ -\ R(1\ -\ e^{-100/23}))}
\label{secondequation}
\end{equation}
where $X$ denotes the exponentially transformed indicator score. The scaling constant (23) ensures approximately 10\% cancellation effect \parencite{mclennanEnglishIndicesDeprivation2019}. In this research, the same scaling constant is applied to adjust rankings, reflecting a moderate level of cancellation and enhancing a more accurate representation of overall liveability. This transformation distributes scores between 0 and 100.

This normalisation approach aligns indicator scores on a common scale while refining aggregation by controlling cancellation effects. This balance between normalisation and aggregation ensures that the CI accurately represents underlying dimensions while considering their interactions. Additionally, it enhances differentiation between neighbourhoods with no amenities (entropy value of -1) and those with only one amenity type (entropy value of 0).
The exponential transformation was applied only after the final scores for each level of indicators and domains within the Liveability Index have been determined.

\subsubsection{Weighting and Aggregation}
While compensability issue during aggregation is addressed through the exponential transformation, another key aspect is determining indicator weights. Weight allocation varies across CIs depending on their objectives. However, this decision can be informed by understanding the intrinsic characteristics of the data and carefully considering the purpose of the CI.

% PCA table (a)
\begin{table*}[h]
\centering
\renewcommand{\arraystretch}{1.3}
\caption{PCA results for diversity and proximity domains}
\label{tab:pca_results}

\begin{subtable}[t]{\textwidth}
\centering
\caption{Diversity Indicators}
\label{tab:diversity_pca}
\small
\begin{tabular}{lccccc}
\toprule
Variable & \multicolumn{5}{c}{Principal Component Loadings} \\
 & \textit{PC1} & \textit{PC2} & \textit{PC3} & \textit{PC4} & \textit{PC5} \\
\midrule
Diversity of Commerce & \textbf{0.506} & -0.078 & -0.250 & 0.354 & \textbf{-0.742} \\
Diversity of Education & 0.426 & -0.192 & \textbf{0.882} & -0.058 & -0.014 \\
Diversity of Entertainment & 0.467 & -0.247 & -0.329 & \textbf{-0.778} & 0.084 \\
Diversity of Healthcare & \textbf{0.501} & -0.100 & -0.221 & 0.498 & \textbf{0.665} \\
Diversity of Living & 0.305 & \textbf{0.941} & 0.050 & -0.134 & 0.029 \\
\midrule
Eigenvalue & 2.882 & 0.817 & 0.572 & 0.424 & 0.305 \\
Proportion of Variance (\%) & 57.6\% & 16.3\% & 11.5\% & 8.5\% & 6.1\% \\
\bottomrule
\end{tabular}
\end{subtable}

\vspace{7pt} % Space between tables

\begin{subtable}[t]{\textwidth}
\centering
\caption{Proximity Indicators}
\label{tab:proximity_pca}
\small
\begin{tabular}{lccccc}
\toprule
Variable & \multicolumn{5}{c}{Principal Component Loadings} \\
 & \textit{PC1} & \textit{PC2} & \textit{PC3} & \textit{PC4} & \textit{PC5} \\
\midrule
Proximity to Commerce & -0.467 & -0.114 & 0.442 & \textbf{0.651} & 0.387 \\
Proximity to Education & -0.397 & \textbf{0.526} & \textbf{-0.672} & 0.066 & 0.332 \\
Proximity to Entertainment & -0.387 & \textbf{-0.711} & -0.477 & 0.115 & -0.324 \\
Proximity to Healthcare & -0.499 & 0.414 & 0.263 & -0.040 & \textbf{-0.713} \\
Proximity to Living & -0.475 & -0.184 & 0.239 & \textbf{-0.747} & 0.355 \\
\midrule
Eigenvalue & 2.360 & 0.829 & 0.775 & 0.613 & 0.423 \\
Proportion of Variance (\%) & 47.2\% & 16.6\% & 15.5\% & 12.2\% & 8.5\% \\
\bottomrule
\end{tabular}

\smallskip
\textit{Note.} PCA loadings exceeding 0.5 (in absolute value) are highlighted in bold.
\end{subtable}
\end{table*}

In this context, Principal Component Analysis (PCA) offers valuable insights. PCA reduces dataset dimensionality while preserving meaningful information by linearly transforming data using the covariance matrix of the standardised dataset, equivalent to correlation matrix  \parencite{jolliffePrincipalComponentAnalysis2016}. This transformation produces a set of uncorrelated principal components, with PCA loadings representing correlation coefficients between these components and original variables \parencite{nardo2005tools}. These loadings quantify how much a variable’s variance can be explained by a principal component, and can inform weight allocation during index construction \parencite{chaoPrincipalComponentbasedWeighted2017}.

Table \ref{tab:pca_results} displays PCA loadings and eigenvalues for diversity and proximity indicators. Selecting components that explain over 90\% of total variance is a common approach \parencite{joint2008handbook}, and in both PCA results, the first four components exceed this threshold. Each indicator is distinctively associated with particular principal components. For instance, diversity of living strongly correlates with PC2 (0.941), diversity of education with PC3 (0.882), diversity of entertainment with PC4 (-0.778), diversity of commerce with PC1 and PC5 (0.506 and -0.742, respectively), and diversity of healthcare with PC1 and PC5 (0.501 and 0.665, respectively). 

Given that each indicator’s highest loadings are distributed across multiple components, it is reasonable to retain all indicators without dimensionality reduction and adopt equal weighting. Equal weighting ensures no single dimension is disproportionately emphasised, a prevalent practice in CIs based on the principle that each variable holds equal importance \parencite{joint2008handbook}. Furthermore, each selected indicator uniquely captures specific facets of liveability. \textcite{nardo2005tools} highlight that equal weighting is particularly useful when correlated indicators provide complementary perspectives rather than redundancy. Considering this research’s aim to holistically depict neighbourhood liveability, the adoption of equal weights for each indicator resonates with its fundamental purpose.

\subsubsection{Robustness Test}
Constructing a CI involves subjective decisions that may introduce uncertainties affecting the final outcomes. Robustness tests assess the CI’s resilience and enhance transparency \parencite{grecoMethodologicalFrameworkComposite2019}. The OECD guidelines identify potential sources of uncertainty, including data selection, normalisation methods, and weighting methods \parencite{joint2008handbook}. To address these, we conducted uncertainty analysis (UA) and sensitivity analysis (SA).

\textbf{Uncertainty Analysis (UA)}
\newline
UA examines how uncertainties in input factors diffuse through the CI, influencing its outputs \parencite{nardo2005tools}. \textcite{saisana2005uncertainty} proposed a Monte Carlo approach, which runs multiple model simulations by randomly selecting input factors $X_i$ (\textit{i} = 1, …, \textit{k}) to analyse the uncertainty’s impact on the index. This involves assigning probability density functions (PDFs) to input factors, generating various combinations of these factors, and subsequently computing the output values for each combination.

This analysis implements a simplified UA, focusing on data selection and normalisation methods in the Liveability Index (LI). Table \ref{tab_5} presents the input factors considered in the simulations. The model runs for \textit{N} = 288 distinct combinations of input factors, denoted as $X^l$, where \textit{l} ranges from 1 to \textit{N}. Each combination is referred to as a sample or a set, represented as $X^l$ = $X_1^l$, \dots, $X_k^l$, with \textit{k} being the total number of input factors. The final LI score and ranking of each neighbourhood were recorded in each simulation.

% Uncertainty Input Factors
\begin{table*}[h]
\centering
\renewcommand{\arraystretch}{1.2} % Increases row height for better readability
\caption{Input factors of the uncertainty analysis}
\label{tab_5}
\resizebox{\textwidth}{!}{ % Ensures table fits within page width
\large % Increases font size
\begin{tabular}{l p{5cm} p{6cm} p{2cm} p{8cm}} % Adjust column widths as needed
\toprule
Input Factor & Name & Interpretation & Range & Values \\
\midrule
X1 & Diversity Indicators & Evaluate the impacts of excluding each individual indicator in the Diversity domain & [1, 6] &
\begin{tabular}[t]{@{}l@{}}
1 = All Diversity domain indicators included \\ 
2 = Diversity of Commerce excluded \\ 
3 = Diversity of Education excluded \\ 
4 = Diversity of Entertainment excluded \\ 
5 = Diversity of Living excluded \\ 
6 = Diversity of Healthcare excluded
\end{tabular} \\
\midrule
X2 & Proximity Indicators & Evaluate the impacts of excluding each individual indicator in the Proximity domain & [1, 6] &
\begin{tabular}[t]{@{}l@{}}
1 = All Proximity domain indicators included \\ 
2 = Proximity to Commerce excluded \\ 
3 = Proximity to Entertainment excluded \\ 
4 = Proximity to Education excluded \\ 
5 = Proximity to Living excluded \\ 
6 = Proximity to Healthcare excluded
\end{tabular} \\
\midrule
X3 & Population Density Indicator & Evaluate the impacts of excluding the Population Density domain & [1, 2] &
\begin{tabular}[t]{@{}l@{}}
1 = Population Density included \\ 
2 = Population Density excluded
\end{tabular} \\
\midrule
X4 & Normalisation Methods & Perform different normalisation methods & [1, 4] &
\begin{tabular}[t]{@{}l@{}}
1 = Exponential transformation used \\ 
2 = MinMax normalisation used \\ 
3 = Robust normalisation used \\ 
4 = Standardisation used
\end{tabular} \\
\bottomrule
\end{tabular}
} % End resizebox
\end{table*}

Suppose CI as the index values for an area \textit{n}, \textit{n} = 1, 2, …, \textit{M} (\textit{M} = 399). The OECD guidelines suggest evaluating the ranking of each area, \textit{Rank}(\textit{$CI_n$}), and the average shift in ranking, $\overline{R}_s$, caused by uncertainties in input factors \parencite{joint2008handbook}. The average change in ranking can be calculated as:

\begin{equation}
\bar{R}_s = \frac{1}{M} \sum_{n=1}^{M} \left| \text{Rank}_{\text{ref}}(CI_n) - \text{Rank}(CI_n) \right|
\label{thirdequation}
\end{equation}
where the reference ranking corresponds to the original index with X1\ =\ 1, X2\ =\ 1, X3\ =\ 1, X4\ =\ 1, and all indicators aggregated equally. Thus, UA examines an area's ranking, \textit{Rank}(\textit{$CI_n$}), while SA analyses the average ranking change, $\overline{R}_s$.

\begin{figure*}[h]
\centering
\includegraphics[width=0.9\textwidth]{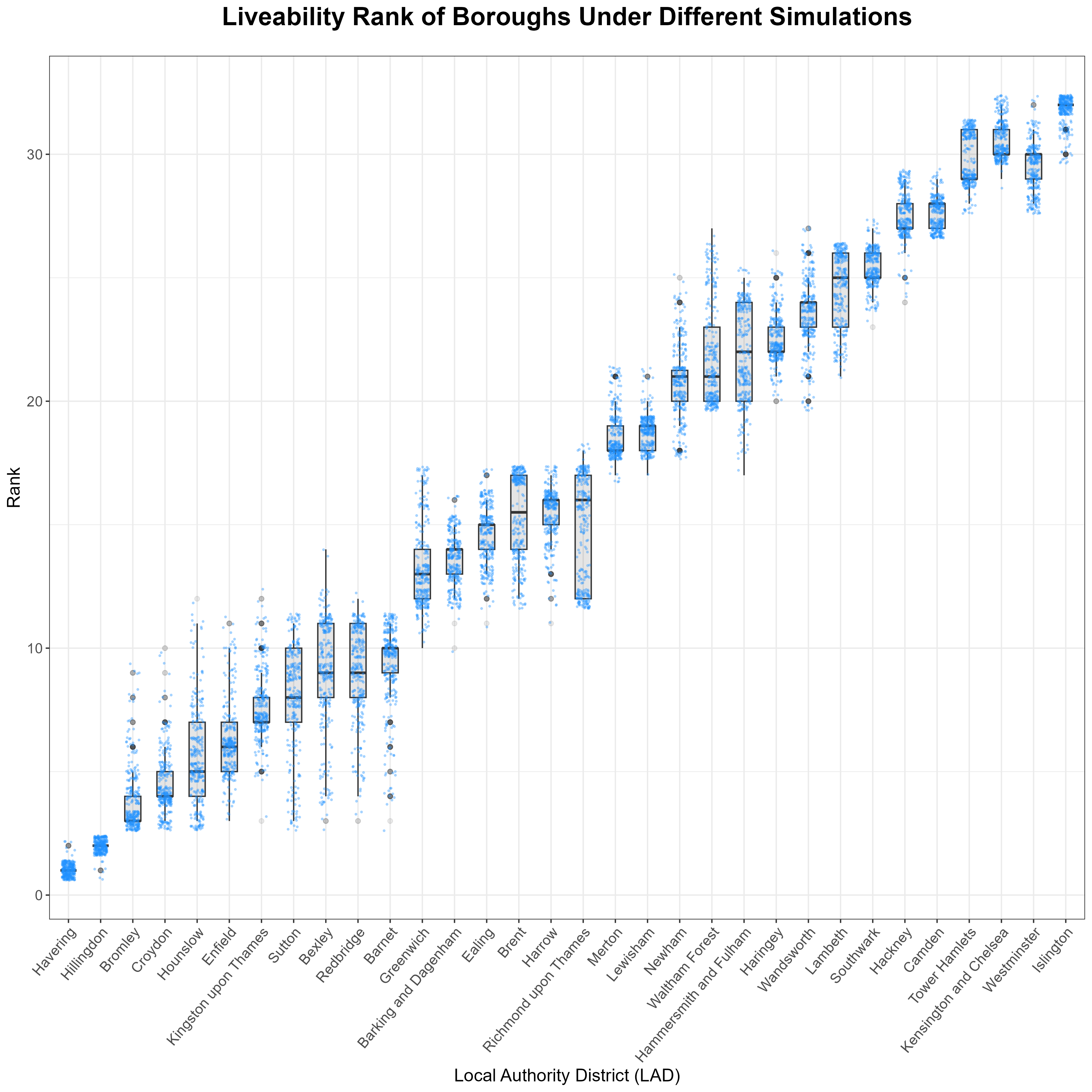}
\caption{Final liveability rankings \textit{Rank}(\textit{$CI_n$}) of boroughs across various simulations. Rankings range from 32 (highest) to 1 (lowest). To simplify interpretation, 399 neighbourhoods are aggregated into 32 Local Authority Districts (LADs). Each data point represents a borough’s ranking under different test configurations.}
\label{ua_fig}
\end{figure*}

In Figure \ref{ua_fig}, borough rankings vary considerably, with greater fluctuation among mid-ranking boroughs and lower variance at the extremes. This may stem from similarities in liveability characteristics among mid-ranked neighbourhoods, making differentiation harder, while larger disparities at the extremes lead to clearer distinctions. Additionally, the exponential normalisation method may amplify small differences in mid-range scores, widening the spread of rankings compared to the extremes.

% UA figure

\textbf{Sensitivity Analysis}
\newline
Sensitivity analysis measures how sensitive the model’s output is to changes in specific indicators. This was conducted by devising the Dirichlet distribution, which models random probability mass functions (PMFs) for a discrete number of indicators. The Dirichlet distribution generates a set of random, non-negative vectors \textit{$p_i$} (\textit{i} = 1, ..., \textit{k}) summing to 1 \parencite{dirichlet}. These vectors represent weights assigned to a set of indicators \textit{$v_i$} (\textit{i} = 1, ..., \textit{k}), with their distribution influenced by a parameter $\alpha$. By varying $\alpha$, we can adjust the weight distribution to emphasise or de-emphasise specific indicators and observe how this affects the model's outcome. The final score, \textit{P}, is computed as a weighted sum of indicators:
\begin{equation}
    \textit{P} = \sum_{i=1}^{k} p_i  v_i 
\label{fourthequation}
\end{equation}
where $p_i$ are Dirichlet-generated sample weights and $v_i$ are the corresponding indicator values.
\newline
The sensitivity analysis was conducted at each aggregation stage: 
\begin{itemize}
    \item Five diversity indicators into the diversity domain score
    \item Five proximity indicators into the proximity domain score
    \item Diversity, proximity, and population density domain scores into the Liveability Index
\end{itemize}

At each stage, 1,000 sample weights were generated for \textit{k} input variables with the initial $\alpha$ value set to 1. The $\alpha$ value of one variable was then gradually increased from 1 to 10, generating 1,000 samples per increment while keeping other variables fixed at 1. This process was repeated for all variables. 
Table \ref{tab_6} summarises the sample weights and $\alpha$ ranges. A total of 50,000 simulations were conducted for diversity and proximity indicators, and 30,000 for domain-level sensitivity analysis.

% sensitivity analysis table
\begin{table}[h]  % use table, not table*, unless you're in a two-column layout
\centering
\renewcommand{\arraystretch}{1.3}
\caption{Number of sample weights used in sensitivity analysis for each aggregation stage}
\label{tab_6}
\small
\resizebox{\textwidth}{!}{
\begin{tabular}{l c c c c}
\toprule
Aggregation Stage & No. Variables Aggregated & No. Sample Weights & Alpha Range & Total Sample Weights \\
\midrule
Diversity Domain & 5 & 1,000 & [1, 10] & 50,000 \\
Proximity Domain & 5 & 1,000 & [1, 10] & 50,000 \\
Liveability Index & 3 & 1,000 & [1, 10] & 30,000 \\
\bottomrule
\end{tabular}
}
\end{table}

\begin{figure*}[h!]
\centering

\begin{subfigure}[t]{0.98\textwidth}
    \centering
    \includegraphics[width=0.8\textwidth, height=0.3\textheight]{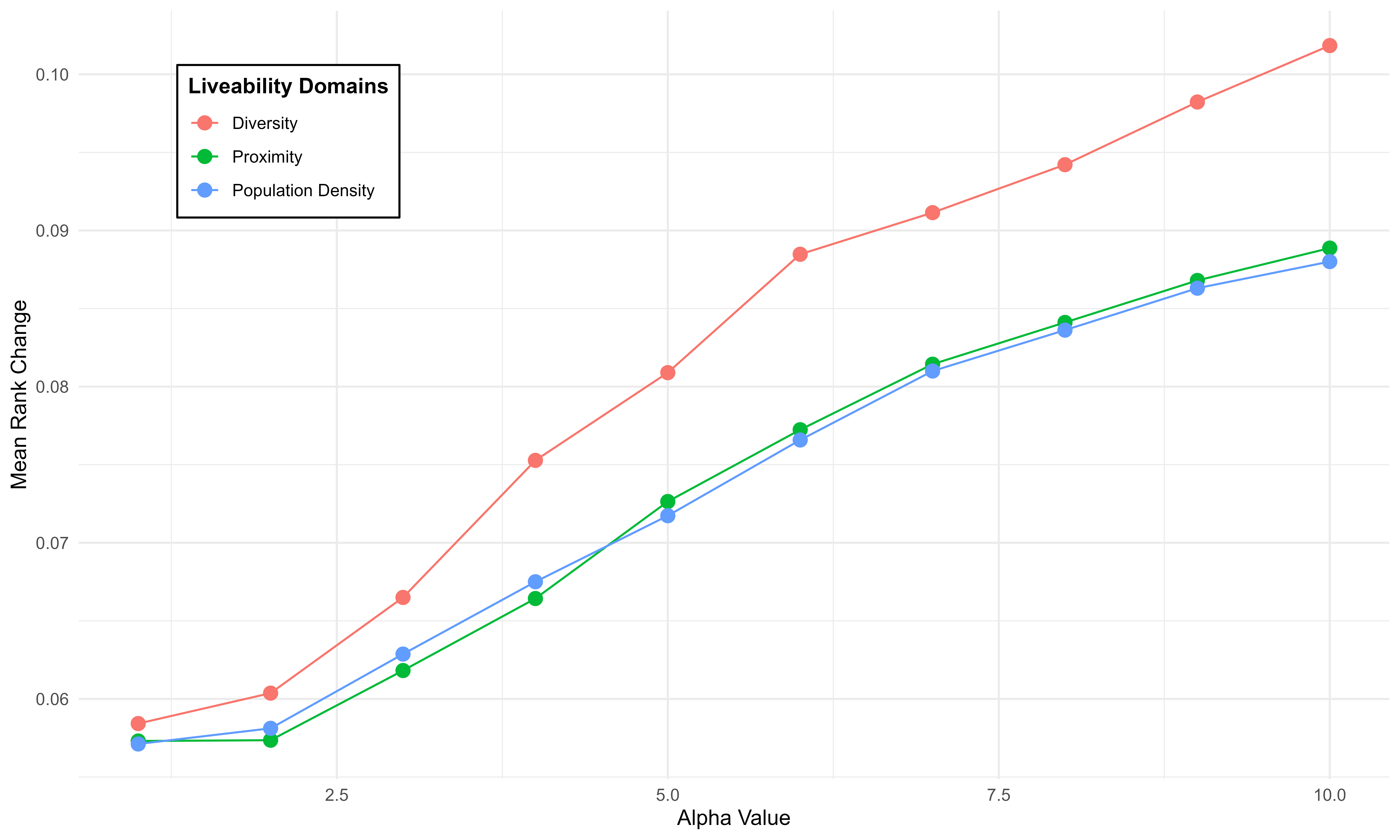}
    \caption{Domain sensitivity analysis}
\end{subfigure}

\vspace{0.5cm}

\begin{subfigure}[t]{0.49\textwidth}
    \centering
    \includegraphics[width=\textwidth, height=0.2\textheight]{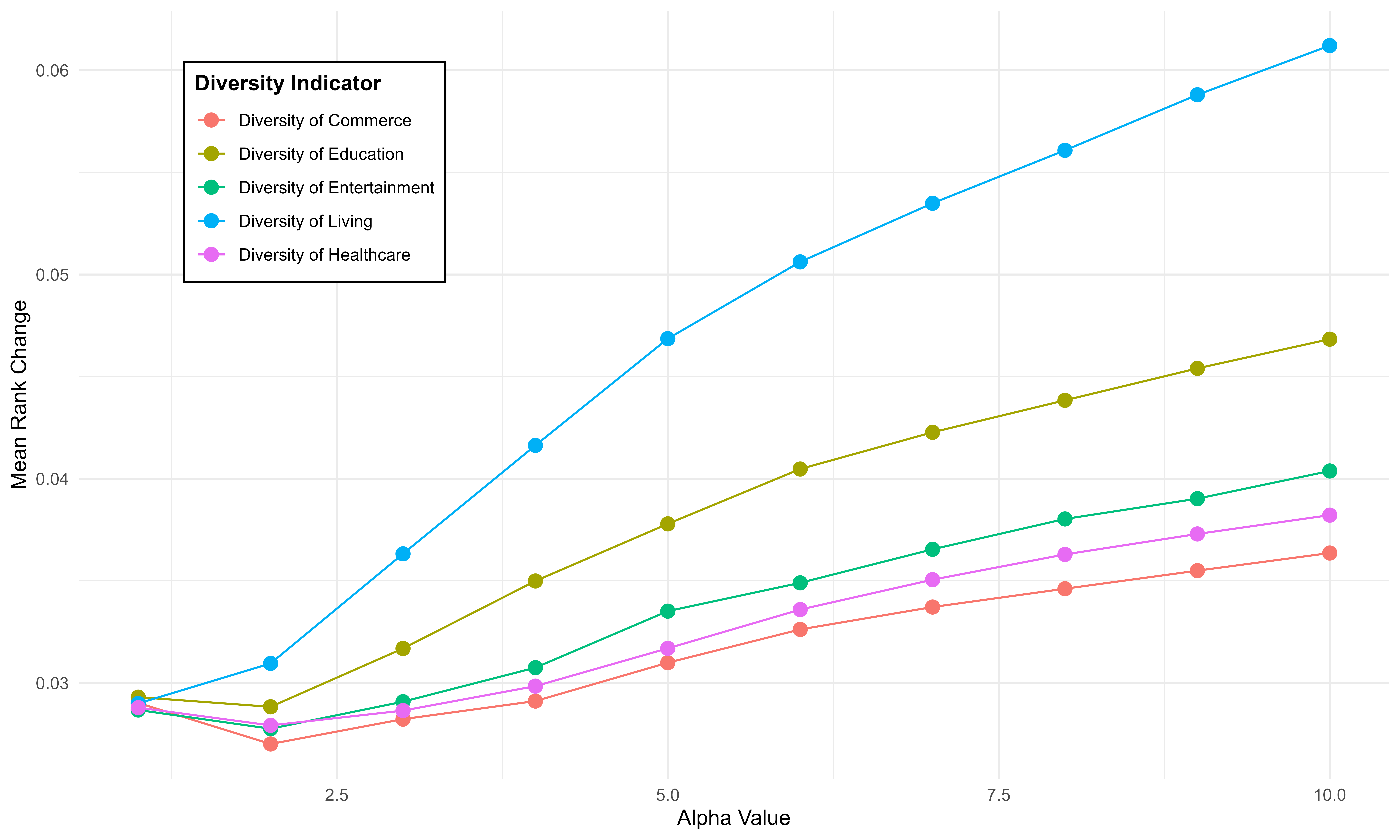}
    \caption{Diversity indicator sensitivity analysis}
\end{subfigure}
\hfill
\begin{subfigure}[t]{0.49\textwidth}
    \centering
    \includegraphics[width=\textwidth, height=0.2\textheight]{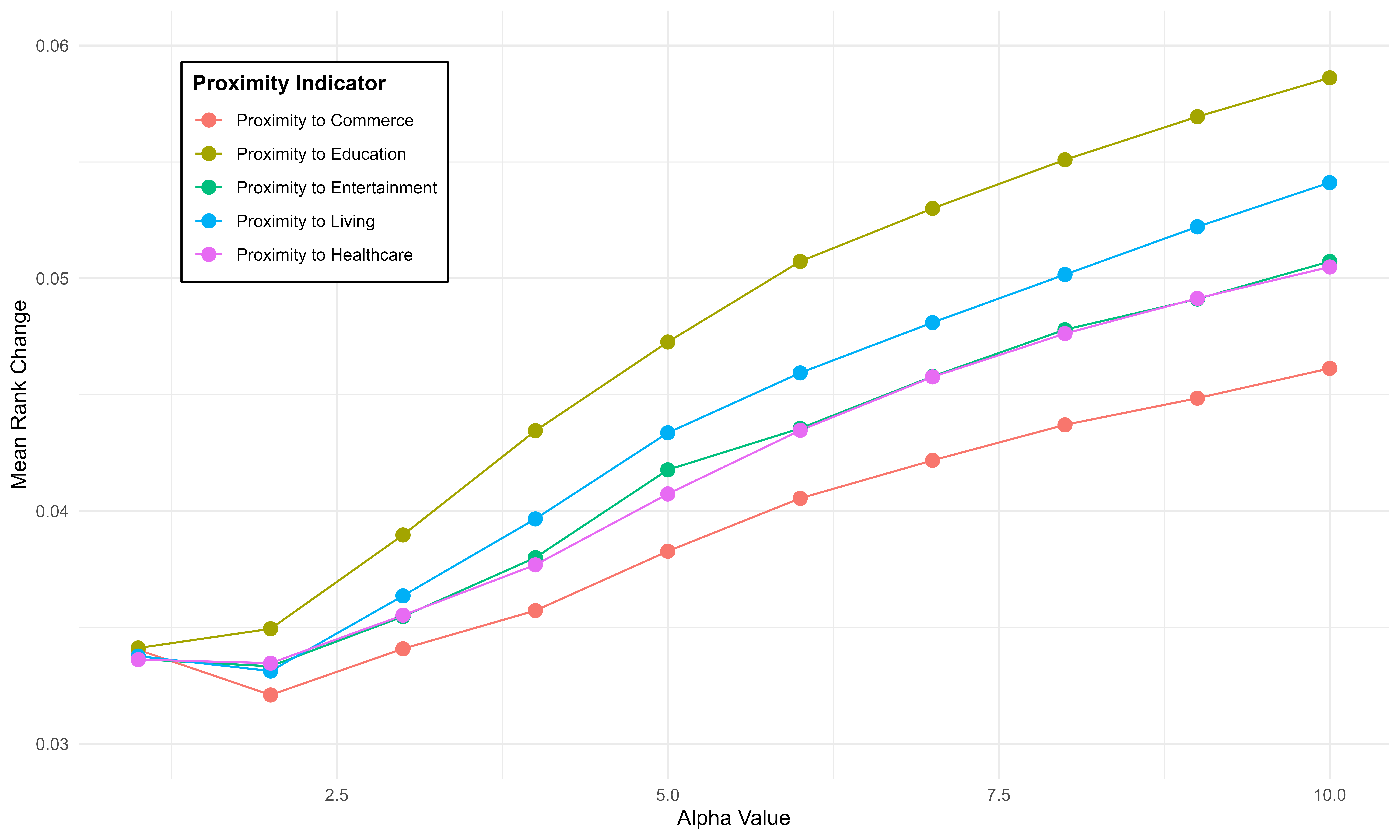}
    \caption{Proximity indicator sensitivity analysis}
\end{subfigure}

\caption{Results of sensitivity analysis: mean $\overline{R}_s$ by $\alpha$ values at each aggregation stage.}
\label{fig_2}
\end{figure*}

The domain-level sensitivity analysis reveals that the diversity domain has a greater effect on mean neighbourhood rank changes than other domains, as illustrated in Figure \ref{fig_2}.a. Its influence becomes pronounced starting at $\alpha$ = 2 and continues to grow across the entire $\alpha$ range. In contrast, while the effects of the proximity and population density domains also grow with increasing $\alpha$, their impacts remain similar to each other and stay below the diversity domain’s influence. Although the scale of $\overline{R}_s$ is smaller at the indicator level, it provides deeper insights into the individual indicators’ contributions. The highest $\overline{R}_s$ values for both indicator-level tests were approximately 0.06, with the diversity of living and proximity to education having the greatest influence, as shown in Figure \ref{fig_2}.b and \ref{fig_2}.c. Proximity indicators generally had a stronger impact, with contributions ranging from 0.45 to 0.06.

% sensitivity analysis figure

\newpage

\subsubsection{Validation}
Validating a newly constructed composite index (CI) against an established index ensures its credibility, reliability, and applicability for decision-making and policy formulation. The English Indices of Deprivation 2019 includes seven indices \parencite{mclennanEnglishIndicesDeprivation2019}, one of which - the ‘Geographical Barriers Sub-domain (GBS)’ within the ‘Barriers to Housing and Services Domain’ – is particularly relevant to the Liveability Index (LI). The GBS measures physical proximity to key local services by assessing how far residents live from essential amenities such as post offices, schools, shops, and GP surgeries. Lower scores indicate higher accessibility. To align with the LI’s geographical scale, the GBS scores were aggregated to the neighbourhood level using a weighted mean approach.

\begin{figure*}[ht]
\centering
\includegraphics[width=0.8\textwidth, height=0.35\textheight]{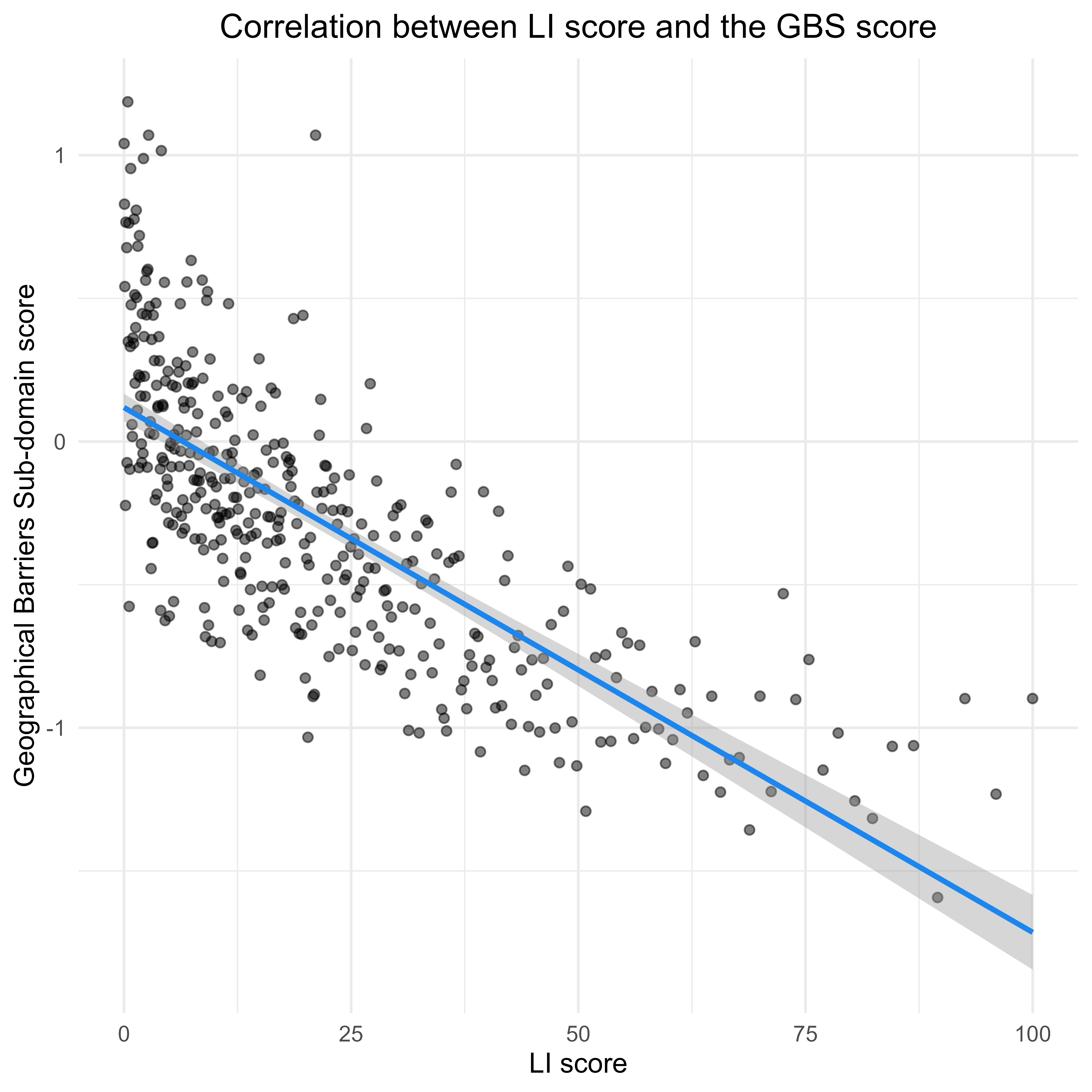}
\caption{Correlation between the LI and the GBS scores at the neighbourhood level. The correlation is statistically significant at the 1\% level, with an adjusted R\textsuperscript{2} of 0.56 and RMSE of 30.07.}
\label{correlation_fig}
\end{figure*}

Figure \ref{correlation_fig} presents the correlation analysis between LI and GBS scores. An adjusted r-squared value of 0.56 suggests that 56\% of the variance in GBS scores is explained by LI scores. The negative correlation indicates that higher liveability corresponds to better proximity. While the correlation highlights similarities in variation, it does not imply causation. Changes in one index do not necessarily drive changes in the other, as additional factors may influence the observed relationship.

% correlation between LI and GBS

\newpage
\subsection{Moran's I Test on GWR Residuals}\label{A3:MoransI}
% Moran's I Spatial Autocorrelation Test Table
\begin{table}[H]
\centering
\renewcommand{\arraystretch}{1.3}
\setlength{\tabcolsep}{12pt} % Adjust spacing as needed
\caption{\centering The results of Moran's I spatial autocorrelation test on GWR residuals}
\label{tab_moran_2}
\small
\begin{tabular}{ccccc}
\toprule
Moran's I & Expectation & Variance & \textit{Z-score} & \textit{P-value} \\
\midrule
0.0043 & -0.0025 & 0.0011 & 0.2011 & 0.4203 \\
\bottomrule
\end{tabular}
\vspace{1pt}
\end{table}

% Residual vs Fitted plot
\begin{figure*}[h]
\centering
\includegraphics[width=0.8\textwidth, height=0.31\textheight]{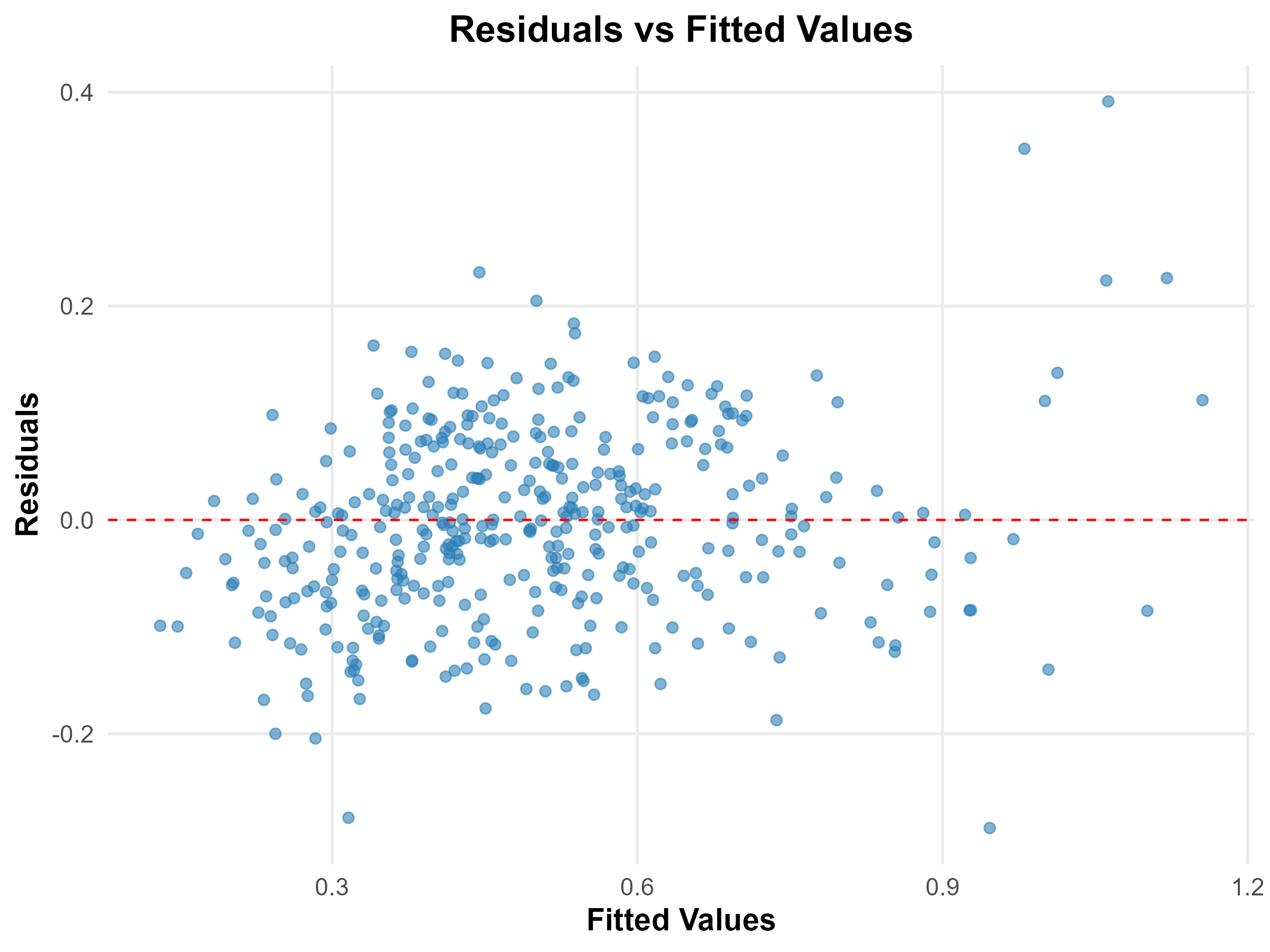}
\caption{Spatial distribution of GWR residuals}
\label{gwr_res_fig}
\end{figure*}

\end{document}